\documentclass{article}

\usepackage{graphicx}
\usepackage{arxiv}
\usepackage[utf8]{inputenc} 
\usepackage[T1]{fontenc}    
\usepackage{hyperref}       
\usepackage{url}            
\usepackage{booktabs}       
\usepackage{amsfonts}       
\usepackage{nicefrac}       
\usepackage{microtype}      
\usepackage{lipsum}		
\usepackage{graphicx}
\usepackage[square,numbers]{natbib}
\usepackage{doi}
\usepackage{gensymb}
\usepackage{xcolor}
\usepackage{rotating}
\usepackage{subcaption}
\usepackage{enumitem}
\usepackage{authblk} 

\title{Centralized Health and Exposomic Resource (C-HER): Analytic and AI-Ready Data for External Exposomic Research}

\renewcommand{\shorttitle}{\textit{arXiv} Template}

\hypersetup{
pdftitle={Spatiotemporal Indexing for Measuring the Spatial and Contextual Exposome},
pdfsubject={Spatiotemporal Indexing},
pdfauthor={Heidi A.~Hanson, Joemy ~Ramsay},
pdfkeywords={First keyword, Second keyword, More},
}

\author[1]{Heidi A. Hanson}
\author[2]{Joemy Ramsay}
\author[1]{Josh Grant}
\author[1]{Maggie Davis}
\author[1]{Janet O. Agbaje}
\author[1]{Dakotah Maguire}
\author[1]{Jeremy Logan}
\author[2]{Marissa Taddie}
\author[1]{Chad Melton}
\author[1]{Midgie MacFarland}
\author[2]{James VanDerslice}

\affil[1]{Oak Ridge National Laboratory, Oak Ridge, TN}
\affil[2]{University of Utah, Salt Lake City, UT}

\begin{document}
\maketitle

\begin{abstract}
The Centralized Health and Exposomic Resource (C-HER) project has identified, profiled, spatially indexed, and stored over 30 external exposomic datasets.
The resulting analytic and AI-ready data (AAIRD) provides a significant opportunity to develop an integrated picture of the exposome for health research.
The exposome is a conceptual framework designed to guide the study of the complex environmental and genetic factors that together shape human health.
Few composite measures of the exposome exist due to the high dimensionality of exposure data, multimodal data sources, and varying spatiotemporal scales.
We develop a data engineering solution that overcomes the challenges of spatio-temporal linkage in this field.
We provide examples of how environmental data can be combined to characterize a region, model air pollution, or provide indicators for cancer research.
The development of AAIRD will allow future studies to use ML and deep learning methods to generate spatial and contextual exposure data for disease prediction.
\end{abstract}


\newpage

\section{Introduction}
The Centralized Health and Exposomic Resource (C-HER) project has identified, profiled, spatially
indexed, and stored 30 external exposomic datasets.
The resulting analytic and AI-ready data (AAIRD) enables the development of an integrated picture of the exposome that can be used for health research.
The exposome provides a conceptual framework for linking the totality of environmental exposures to human health outcomes---from conception to death  \cite{wild2012exposome}.
When translating this abstract concept into concrete measurable values, the field of exposomics is still in its infancy.
Although some advances have been made in measuring the internal exposome \cite{david2021towards}, measuring the external exposome is still severely limited by significant methodological and engineering challenges \cite{zheng2020design,hu2023methodological}.
Methodologically, quantifying the contribution of environmental exposures across space and time requires analytic approaches that can deal with high-dimensional, correlated, and multi-fidelity data.
Recent advances in AI have the potential to yield new solutions to these challenges, but this promise cannot be realized without large volumes of interoperable data.
Datasets for measuring external environmental exposures are multimodal in nature, recording information from sources such as environmental monitoring stations, satellite imagery, geological surveys, and demographic and national health surveys.
Although these datasets are often publicly available, they are not interoperable.
The diverse datasets are generally collected by the scientific community for purposes other than exposure measurement, resulting in data that vary in format, quality, spatiotemporal resolution, and spatial extent.
For a single project, the required data cleaning steps often lead to a huge data management overhead that requires highly technical teams and significant time and computational costs, making external exposomic research beyond reach for many.
Therefore, reproducible data pipelines to create AAIRD are needed to operationalize the external environment.

\paragraph{Harmonizing Heterogeneous Data Formats.} Exposomic data are stored in multiple formats, including raster, tabular, and vector data.
Computational workflows for reproducible extraction, transformation, and loading (ETL) of data are necessary to ensure high-quality, well-documented, and accessible external exposomic data.
ETL processes for heterogeneous datasets vary in complexity.
Tabular data are often easy to load and have high cross-platform portability but require additional steps to create geometry or time fields for spatial analyses.
Datasets (e.g., satellite imagery) require more complex geoprocessing tools that can handle hierarchical data to process and store raster and image data.
For these reasons, a common code repository or data catalog that provides links to external datasets is helpful, but it stops short of removing the barriers associated with data cleaning overhead and solving the harmonization requirements needed to create measures of the external exposome.

\paragraph{Harmonizing Data with a Common Spatial Extent.} Once data have been loaded into a centralized database, transforming the data into a uniform grid (e.g., Uber H3 hexagonal grid system) makes it possible to measure features on a common spatial scale.
Although conceptually intuitive, the sheer volume, heterogeneity, and complexity of exposomic datasets present significant challenges for workflow development.
Converting exposomic measures collected at different spatial scales requires complex workflows for spatial interpolation and change of support when ensuring data quality and consistency at the population scale.
Adding to the complexity is the need for context and data-type specific decisions to ensure accurate spatial interpolation.
Creating flexible workflows for spatially indexing data that can be easily tailored for specific use cases will lower the barrier of entry for external exposomic research.

\paragraph{Analysis and AI-Ready Spatiotemporal Data.} The most useful data for causal inference in exposomic research is measured at small spatial and temporal resolutions, which leads to huge storage and memory requirements.
Solutions for mitigating data issues (e.g., systematic missingness, data sparsity) often require sophisticated modeling techniques for spatial interpolation and downscaling (small-area estimation).
The most accepted methods for spatial interpolation are computationally expensive ensemble ML methods that use a broad range of data---from areal imagery to ground-level sensors.
These methods have proven useful in exposure modeling, but few researchers have the computational capabilities necessary to implement and scale these models to the population level.
The creation of multi-array, multi-thematic, analysis-ready collections of data could serve as a foundation for building comprehensive models of the exposome, lowering data access barriers, and facilitating external exposomic research.

\paragraph{Linking Spatiotemporal Data to Longitudinal Residential Histories for Health Research.} The development of chronic diseases (e.g., cancer) can take years or even decades, so exposomics data must correspond to the long temporal window during which an individual's risk factors are accumulated.
This can be accomplished by linking exposure measures with longitudinal residential history data, enabling the capture of changing environmental conditions, neighborhood characteristics, and exposure levels.
However, residential history data are inherently sensitive because an individual's residential trajectory could be used to identify them.
However, because AAIRD is spatially indexed, it can be shared with external institutions to facilitate exposomic research with standardized datasets while ensuring privacy protection.
This indexing allows exposure data to be securely transferred to institutions or researchers who can then efficiently link the data by using the associated hexagonal identifier.

Although there are some centralized and federated data ecosystems for external exposomic data (e.g., Harvard Dataverse \cite{jiaoharvard}), these tools provide centralized access to data and educational tools, but they do not provide solutions for AAIRD.
Solutions are needed to process diverse data in a way that is easy to query and link across space and time while maintaining flexibility such that it can be applied across studies and health outcomes.

In this paper, we outline an AAIRD ecosystem for exposomic data that can be used to bridge this gap.
We describe (i)~spatial indexing for exposomic data, (ii)~an example of air pollution modeling with spatially indexed data, (iii)~examples of unsupervised clustering with spatially indexed data, and (iv)~how these data are being linked to residential histories for cancer research.
The primary goal is to show how exposomic AAIRD can make it easier to create holistic measures of the exposome that can be used across scientific disciplines.

\section{Materials and Methods}
\subsection{Design of the Centralized Health and Exposomic Resource (C-HER) Database}

\paragraph{Reproducible Workflows for Exposomic Data.} The C-HER database enables researchers to develop measures of environmental context (external exposome) for individuals over space and time.
We utilize open-source software such as S3, PostgreSQL, Prefect, and Python to develop workflows for loading, transforming, and managing data within the C-HER framework.
Prefect is an effective workflow orchestration tool that serves as the backbone of our ETL system.
It allows us to design modular and repeatable tasks that can be reused across multiple workflows.
These tasks can then be coordinated using the flow management tool to ensure a logical sequence of tasks for each workflow.
MinIO serves as the primary storage system for raw and untransformed data in our pipeline.
This open-source and high-performance object storage system is well suited for large geospatial datasets.
PostgreSQL is used to store tabular data and manage metadata within our ETL system.
We have enabled strict role-based access control mechanisms to ensure that only authorized users and systems can access data.
In the future, we plan to open source the tools and share these code bases to enable validation of results and to follow findable, accessible, interoperable, and reusable (FAIR) principles.
The C-HER repository stores five types of data: 

\begin{enumerate}[label=(\roman*),leftmargin=*]
    \item \textit{Raster Data}: Satellite images, aerial photographs, and derived raster products.
    These files are stored in formats such as GeoTIFF, HDF5, NetCDF, and GRIB.
    \item \textit{Vector Data}: Geographic features such as boundaries, roads, and points of interest.
    These data are stored and managed in SQL format within PostgreSQL databases.
    \item  \textit{Model Data}: Processed model output and derivative data that are not trivial to recreate and can be reused to measure the exposome.
    This includes data generated by complex models and/or data that require specialized hardware to recreate.
    \item \textit{Data Ingestion Code}: Links to the code used to ingest, transform, and create data are stored in the database.
\end{enumerate}

\subsection{Spatially Indexing External Exposome Data}

Easily aligning measures of the external environment across varying levels of spatial and temporal aggregation is one of the significant challenges that prohibits the development of comprehensive measures of the external exposome.
A common spatial indexing scheme must be selected to allow us to easily construct measures of the exposome.

\paragraph{Conversion to Standardized Hexagonal Grid.} Spatially linked exposure data are increasingly important for exposomic research.
We migrated area-level, demographic, and environmental data into a spatially harmonized architecture.
Grid systems are critical for analyzing, visualizing, and exploring spatial data.
Hexagonal grid cells minimize quantization error with movement and allow for easy approximation of radiuses.
We used the \href{https://www.uber.com/blog/h3/}{Uber H3} open-source hierarchical hexagonal grid indexing system to spatially harmonize data.
H3 supports 16 resolutions, and each finer resolution is generated by subdividing the coarser resolution cell into 7 child cells.
Cell areas range from 4,357,449~km$^2$ at the coarsest resolution to 0.000000895~km$^2$ at the finest.
Movement between resolutions is also efficient because of the hierarchical nature of the grid cell indices.
H3 serves as a bridge between vector and raster data by acting as a vectorized representation of raster data, or alternatively, a raster framework enhanced with vector-like capabilities.
By placing exposure datasets into this grid system, we can efficiently combine data sources to generate multiple exposure measures as well as improve data visualization/mapping, storage, and shareability.
This approach also poses significant challenges, which are described in detail in~\hyperref[sec:supplement-1]{Supplement 1}.

All exposure datasets were converted to Uber H3 level-8 resolution, hex8, which has hexagons with an average area of 0.737~km$^2$ and average edge length of 0.53~km.
The point location or native grid for each exposure dataset was overlaid with the hexagonal grid to obtain the proportion of the native grid cell contained within a hex8 grid cell.
These proportions were then used to convert exposure values from the native grid to the hex8 grid using a spatially weighted average.

\subsection{Low-Resource and Near Real-Time PM2.5 Modeling}
Data from the C-HER database are combined with the attention-based Senseiver framework~\cite{santos2023development} to estimate average daily surface-level concentrations of PM2.5 (fine particulate matter 2.5~{\textmu}m or smaller) in the Ohio River Valley in 2018.
A more detailed description of this model can be found in the literature.
In short, data from multiple sources that had been spatially indexed at the hex8 level were extracted from C-HER and used as static or time-varying features in the Senseiver model.
The AAIRD and the low-resource Senseiver model can be used to generate near real-time PM2.5 predictions---a feat that has not been accomplished for epidemiological research.
Data from the following sources were extracted from C-HER and used to train the model:
\begin{enumerate}[label=(\roman*),leftmargin=*]
    \item Global Multiresolution Terrain Elevation Data 2010~\cite{Elevation} provided by the United States Geological Survey (USGS) at a resolution of 30 arc seconds.
    \item Seven meteorological variables from the GridMet database, which provides daily weather data with a resolution of 4~km.
    \item The 2013 National Land Cover Database from the USGS.
    The population density estimates from the 2016 Landscan data are derived from raw data at a 3 arc-second ($\sim$90~m) resolution.
    \item PM2.5 concentration data from the US Environmental Protection Agency's (EPA's) Air Quality System \cite{epa_aqs_faqsd}.
\end{enumerate}

\subsection{Exposure Datasets}

Although the complete C-HER data repository includes more than 30 exposomic datasets, this demonstration analysis incorporated a subset of high spatial and temporal resolution contextual information, demographic, and environmental exposures, including criteria air pollutants, air toxics, radiometric data, and elemental carbon.
We use descriptive and thematic mapping to highlight the utility of AAIRD for multiple use cases.

\paragraph{Criteria Air Pollutants.} Daily estimates of exposure to PM2.5 were obtained from the Fused Air Quality Surface Using Downscaling (FAQSD) model published by the EPA \cite{epa_rsig_2015}.
FAQSD is a Bayesian space-time downscaler model that combines daily 24-hour average PM2.5 monitoring data from the National Air Monitoring Stations/State and Local Air Monitoring Stations with 12~km gridded output from the Models-3/Community Multiscale Air Quality model.
Daily predictions are generated at the 2010 US Census Tract level.

\paragraph{Wildfire Smoke.} Daily estimates of smoke PM2.5 for the contiguous US in 2020 were downloaded at a 10~km resolution \cite{childs2022daily}.
Smoke PM2.5 estimates were generated for days when satellite imagery--based plume classification and simulated air trajectories originating at fires indicated the presence of smoke.
A ground-based measure of smoke PM2.5 was generated for these days by modeling deviations from recent location- and month-specific median non-smoke PM2.5 at EPA monitoring stations.
Daily estimates of PM2.5 attributable to wildfire smoke were then modeled at a 10~km resolution over the contiguous US.

\paragraph{Air Toxics.} Exposure to carcinogens from industrial sources for 2018 was estimated using the Risk-Screening Environmental Indicators Geographic Microdata (RSEI-GM) model results, a screening-level tool used to compare chemical levels and trends around reporting facilities.
RSEI-GM is published by the EPA and provides granular estimates (810~$\times$~810~m grid) of air concentrations within 50~km of each reporting facility.
Air concentrations for over 700 chemicals are reported using models that incorporate information on chemical release levels from the Toxics Release Inventory (TRI), facility-specific information, and chemical fate and transport.
TRI reporting is required for industrial and federal facilities in the US in specified industrial sectors, including manufacturing, metal mining, power generation, and hazardous waste treatment.
The model concentrations were converted from the annual hourly average concentration to the annual daily average concentration for each grid cell.
Carcinogens were identified from agents classified by the International Agency for Research on Cancer (IARC) Monographs, Volumes 1--135; National Toxicology Program (NTP) 15th Report on Carcinogens; and EPA’s Integrated Risk Information System (IRIS) program.
All chemicals classified as \textit{Carcinogenic to Humans} (IARC Group 1, IRIS Group A, and NTP \textit{Known}) or \textit{Probably/Possibly Carcinogenic to Humans} (IARC Group 2A/B, IRIS Group B1/2, and NTP \textit{RAHC}) were included for linkage with RSEI-GM using a CAS number for a total of 268 carcinogens linked.
Diagnosis sites were also extracted from text strings contained in the IRIS report, and carcinogens associated with cancer at the following sites were identified: bladder, breast, kidney, liver, lung, lymphoma, sarcoma, and thyroid.

\paragraph{Radiometric Data.} Data from the National Uranium Resource Evaluation program was used to measure equivalent uranium (ppm), equivalent thorium (ppm), and potassium (ppm) across the study area \cite{hill2009aeromagnetic}.
Solar radiation, or daily total shortwave radiation, measures from 2019 DayMet data were averaged across the year to construct an average measure of radiation exposure \cite{thornton2022daymet}.
The EPA map of radon zones was used to estimate the indoor radon exposure \cite{epa1993radon}.
Data were downscaled to the hex level by assigning the county-level value to all intersecting hexes.
For cases in which hexes overlay county boundaries, the spatial area of the intersect is used to assign the most prominent category for the area.

\paragraph{Elemental Carbon.} Annual estimates of PM2.5 elemental carbon were downloaded for 2006--2019 for the contiguous US \cite{amini2022hyperlocal}.
Concentrations were modeled at two spatial resolutions: a 50~$\times$50~m hyper-resolution in urban areas and a 1~$\times$~1~km high-resolution in non-urban areas.
The annual PM2.5 composition data was generated using an ensemble approach that combines ML models with either a generalized additive model ensemble geographic weighted average or a super-learning approach.
From this, 3,535 urban areas were identified across the US for modeling at the hyper-resolution.
The rest of the country was classified as non-urban.
The models were validated using 10$\times$ cross-validation, yielding overall R$^2$ values ranging from 0.910 to 0.970 on the training sets and from 0.860 to 0.960 on test sets.

\paragraph{Population Masking.} Approximately 47\% of the US is uninhabited.
As a result, population masks can be used to reduce the volume of data for use cases in which contextual area--based measures are needed only for inhabited areas.
The Landscan population density \cite{bhaduri2002landscan} was used to generate the population masks we used to demonstrate the utility of this method.
Landscan data uses satellite imagery to map populated areas around the world at a 30~m resolution.
Any hex8 grid cell with at least one person living in it was included in the mask, and uninhabited cells were excluded.

\subsection{The Cancer Outcomes and Neighborhood Environmental Correlations Tracker (CONNECT) Database}
The National Cancer Institute's Surveillance, Epidemiology, and End Results (SEER) program linked cancer patients to LexisNexis residential history data across 11 SEER registries \cite{stinchcomb2016nci, tatalovich2022assessment}.
LexisNexis maintains a database of residential history on more than 276 million US individuals \cite{LexID}.
In this paper, we briefly describe the feasibility of linking residential location data between 1995 and 2024 from four SEER registries to air pollution, indoor radon, and the RSEI-GM data.
C-HER data was extracted and used to create a relational database that can be used by SEER to process, link, and extract environmental exposure data for cancer patients.
The Cancer Outcomes and Neighborhood Environmental Correlations Tracker (CONNECT) enables easy and privacy-preserving retrieval of exposomic data for downstream analyses.

\subsection{Thematic Mapping}
The relationship between single exposures (e.g., PM2.5) and health outcomes has been well documented.
However, it is rare for individuals to be exposed to a single pollutant in isolation \cite{kodros2024cumulative, molitor2011identifying}.
We provide two use cases that highlight how spatially indexed data can be used to identify high-risk zones.
First, we map the co-occurrence of PM2.5 and wildfire smoke in September 2000.
Second, we develop a new metric to quantify the spatial distribution of air toxics classified as carcinogenic to humans (Group 1), probably carcinogenic to humans (Group 2A), or possibly carcinogenic to humans (Group 2B).
Using the RSEI-GM, we calculate a cumulative excess exposure mixture (CEEM) score to quantify carcinogenic air toxic exposures in 2000 for each hex8 grid cell.
The CEEM score was calculated for each hex using the following equation:

\begin{equation}
CEEM_j = (\sum_{i=1}^{n} C_i\div L_i),
\end{equation}

where \(CEEM_j\) is the equivalent cumulative exposure score from the mixture of chemicals, \(C_i\) is the concentration of carcinogen \textit{i}, and \(L_i\) is the exposure limit for carcinogen \textit{i}.
Exposure limits were defined using the inhalation reference concentration from IRIS.
An \(E_M > 1\) indicates that exposure limits are exceeded at the mixture level, even if individual carcinogens are below their limits.

\subsection{Unsupervised Clustering}
The integration of datasets from multiple environmental sources can be used to identify vulnerable populations and link the exposome to a wide range of health outcomes.
We group C-HER datasets into different thematic mapping applications and perform clustering analyses using the multimodal exposomic data.
For demonstration, we selected the states in the Ohio River Watershed: Illinois, Indiana, Ohio, Pennsylvania, Kentucky, and West Virginia.
We highlight three use cases: (i)~uncovering temporal trends in exposure to elemental carbon; (ii)~identifying spatial patterns of exposure to sources of low-dose radiation; and (iii)~identifying exposomic signatures using elemental carbon, radiometric data, PM2.5, and toxic releases.
The Hierarchical Density Based Spatial Clustering of Applications with Noise (HDBSCAN) framework was used for all three use cases because we sought an unsupervised data-driven approach that requires a limited amount of hyperparameter tuning.
HDBSCAN can effectively identify spatial clusters with arbitrary shapes, handle clusters with varying densities, and filter out noise \cite{campello2013density, campello2015hierarchical}.
Hyperparameters were selected using a grid search approach in which the minimum sample and cluster sizes were allowed to vary from 5 to 100.
The best fitting model was selected using Silhouette scores.
We describe the different methods used for these use cases below.

\paragraph{Elemental Carbon.} A single temporal exposure vector was generated per hex, with each column representing the annual elemental carbon concentration for each year from 2006 to 2019.
All values were standardized prior to clustering using HDBSCAN, with a minimum cluster size of 50 and a minimum sample size of 60.
No spatial data were included in this analysis, and the aspatial approach was selected to allow for identification of global temporal exposure patterns.
Following the HDBSCAN analysis, elemental carbon distributions were summarized for each cluster.

\paragraph{Radiometric Exposure.} Universal Transverse Mercator codes were included as features in the model to incorporate the spatial structure of the data and capture local patterns of exposure.
Radiometric data embeddings were then created with principal component analysis (PCA).
The first three principal components (PCs), which together accounted for 90\% of the total variance, were retained for subsequent cluster analyses and thematic mapping.
All features were standardized using the \texttt{StandardScaler} function in Python.
Following the HDBSCAN analysis, we summarized the radiometric measures for each cluster using a radar chart.
The features were normalized on a 0--10 scale to improve visualization.

\paragraph{Carcinogenic Exposure.} For the cluster analysis, we chose to mask all unpopulated areas because our goal was to show populated areas with high concentrations of chemicals known to have strong associations with cancer risk in humans.
RSEI chemical exposure data from 2018 were used to estimate the amount of carcinogenic exposure per chemical in each hex.
For this analysis, we used a definition of \textit{carcinogenic} that expanded the list of chemicals to any chemical that was classified as carcinogenic in the EPA's RSEI risk calculations, regardless of the weight of evidence~\cite{epa_rsei_methodology_2023}.
A total of 102 carcinogenic chemicals were released in the Ohio River Valley in 2018, which
resulted in exposure measures for these 102 unique chemicals.
The features were standardized, and the carcinogen exposure embeddings were created with PCA.
We selected the top 25 PCs based on the elbow method.
We used an aspatial approach for this analysis, allowing us to identify global patterns in carcinogenic exposure regardless of geographic location.

\paragraph{Combined Analysis.} We used a subset of the exposure datasets mentioned above to identify exposomic signatures in the 2018 dataset.
Elemental carbon, radiometric data, PM2.5 estimates from the Senseiver model, and carcinogenic PCs were used for this analysis.
Wildfire smoke was not used because the levels of surface-level smoke PM2.5 were low in this region in 2018.
All features were standardized using the \texttt{StandardScaler} function in Python.
Following the HDBSCAN analysis, we summarized the measures for each cluster using a radar chart.
The features were normalized on a 0--10 scale to improve visualization.

\section{Results}

\subsection{AAIRD for Near Real-Time PM2.5 Modeling}
\begin{figure}[h]
    \begin{subfigure}[b]{0.35\textwidth}
        \centering
        \includegraphics[width=\linewidth]{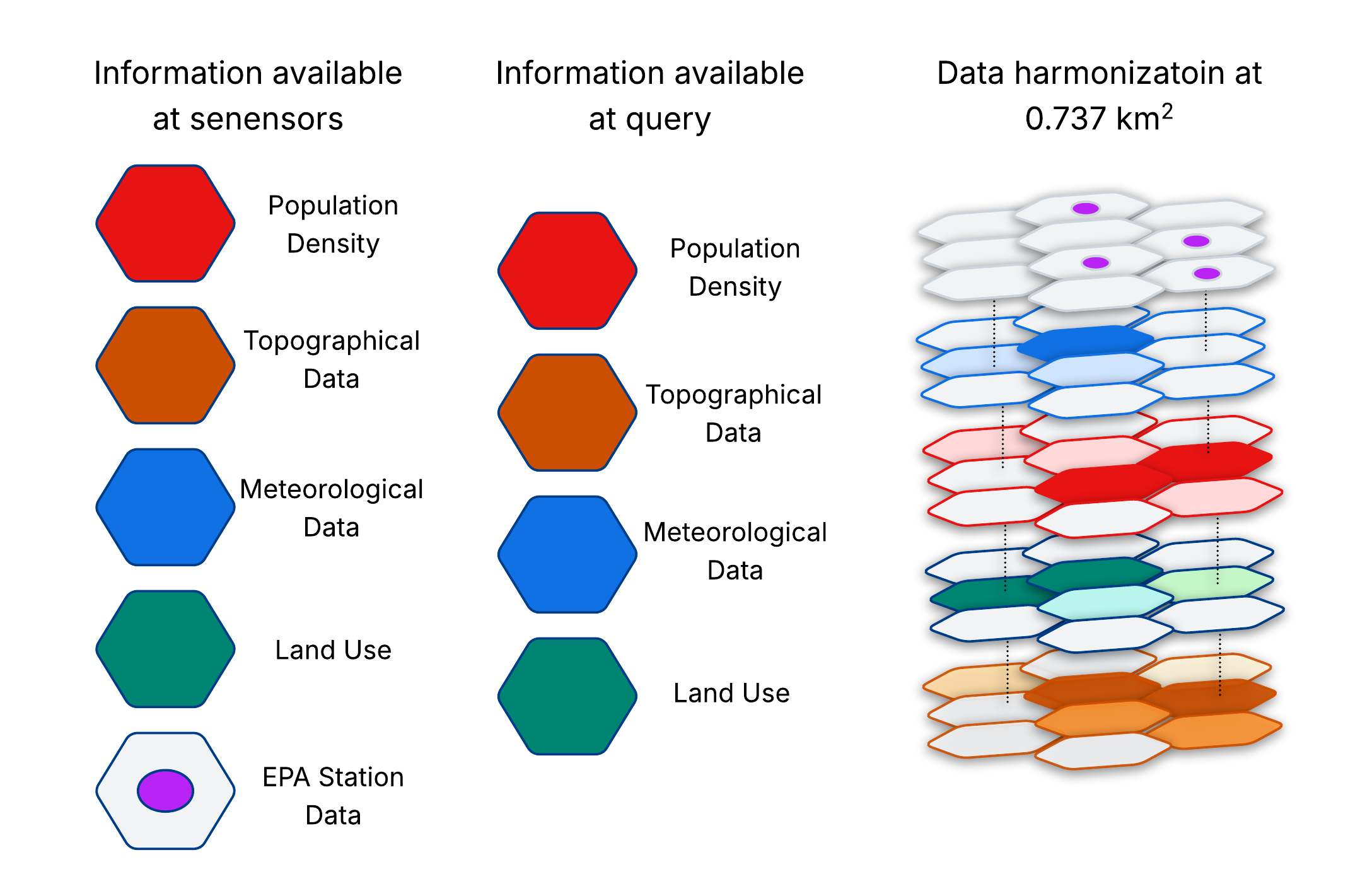}
        \caption{Average daily PM2.5 exposure modeled with Senseiver.}
        \label{fig:AP_Model}          
    \end{subfigure}
    \hfill
    \begin{subfigure}[b]{0.60\textwidth}
        \centering
        \includegraphics[width=\linewidth]{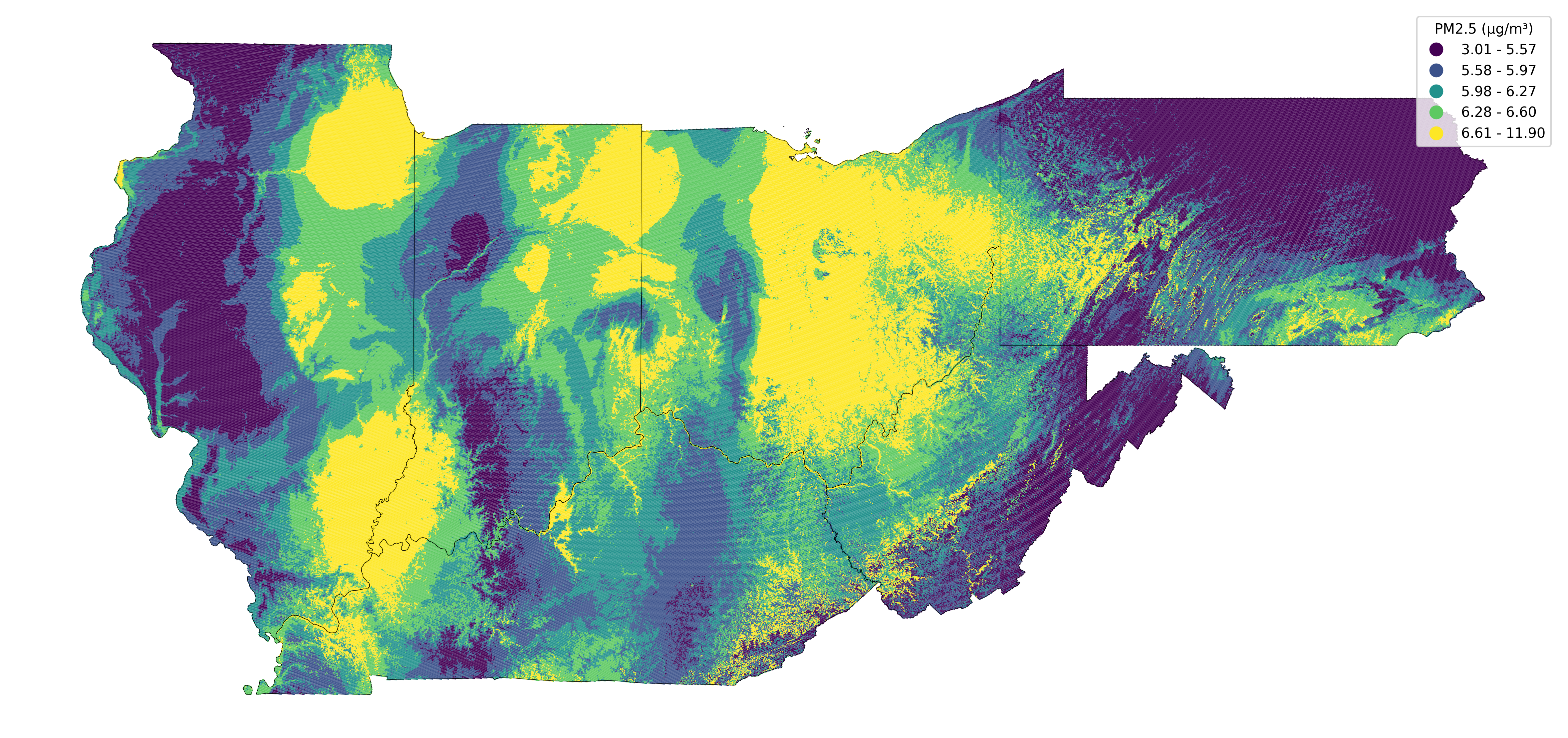}
        \caption{Quintiles of average daily PM2.5 exposure in 2018 modeled with Senseiver.}
        \label{fig:SensevierResults}       
    \end{subfigure}
    \caption{Ground-level, real-time modeling for PM2.5 pollution predictions.
(a) Schematic showing the benefit of spatially indexing data for downstream modeling applications.
(b) Daily average PM2.5 exposure in 2018 modeled with C-HER data and the Senseiver architecture.}
    \label{fig:Sensevier}
\end{figure}

Preprocessed data and AAIRD enable near real-time PM2.5 modeling at scale.
Population density, topographical data, and land use data are all static datasets when the air pollution prediction timescale is days or weeks.
C-HER facilitates easily accessible data that can be quickly merged using the hex8 identifier.
C-HER's reproducible workflows provide low-code solutions for incorporating time-varying data, such as PM2.5 values at monitoring stations and meteorological data.
Figure \ref{fig:Sensevier} shows the quintiles of daily PM2.5 in 2018 estimated with Senseiver.
During this time period, the National Ambient Air Quality Standard for PM2.5 was 12~\textmu{g}/m$^3$, and all areas fall within the attainment level.
The highest levels for this period are clustered around densely populated areas in Illinois, Indiana, and Ohio.
 
\subsection{AAIRD for Thematic Mapping}
\begin{figure}[h]
    \centering
    \begin{subfigure}{0.45\textwidth}
        \centering
        \includegraphics[width=\linewidth]{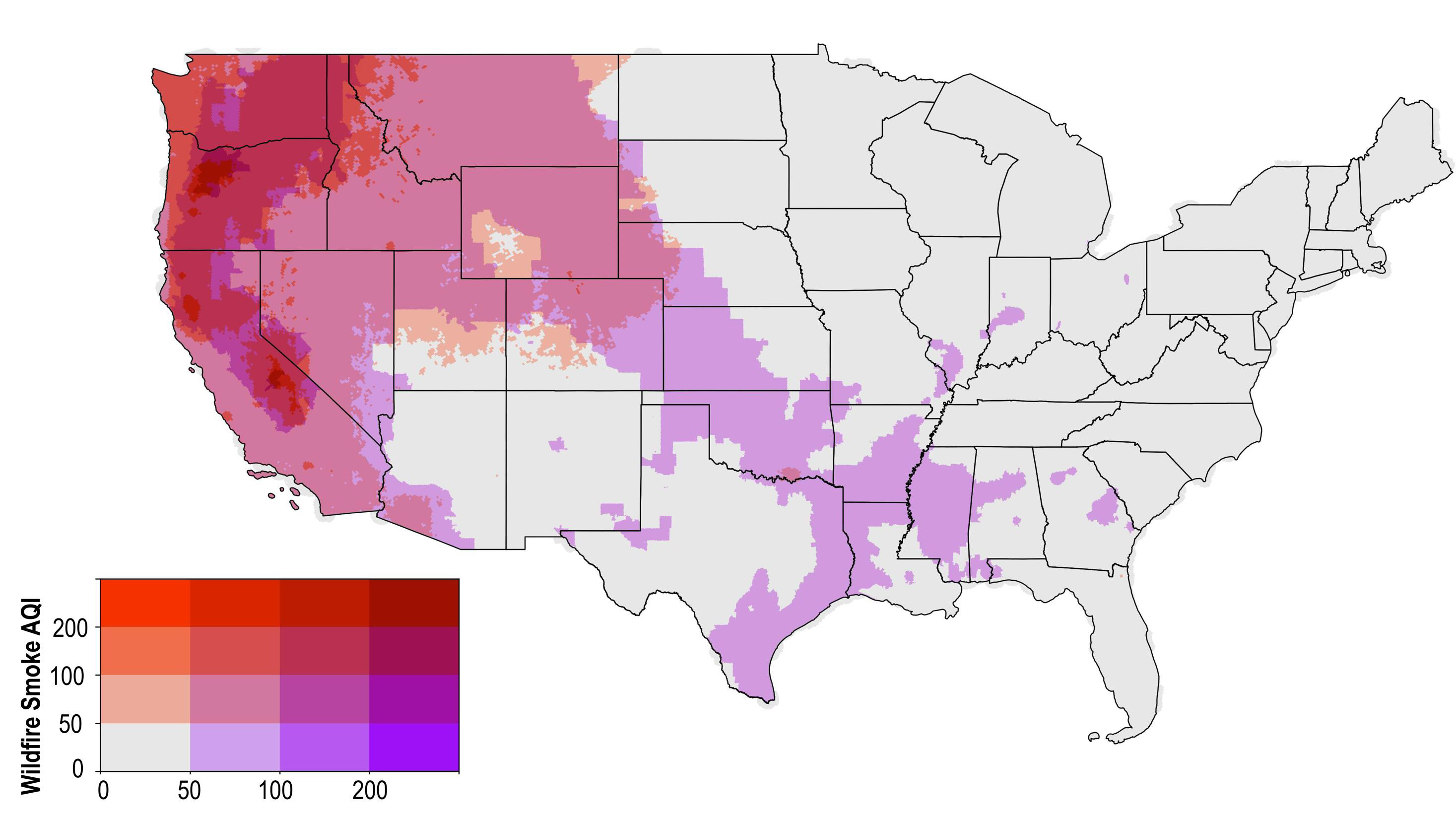}
        \caption{Bivariate mapping of wildfire smoke and PM2.5.}
        \label{fig:smoke_map}
    \end{subfigure}
    \hfill
    \begin{subfigure}{0.53\textwidth}
        \centering
        \includegraphics[width=\linewidth]{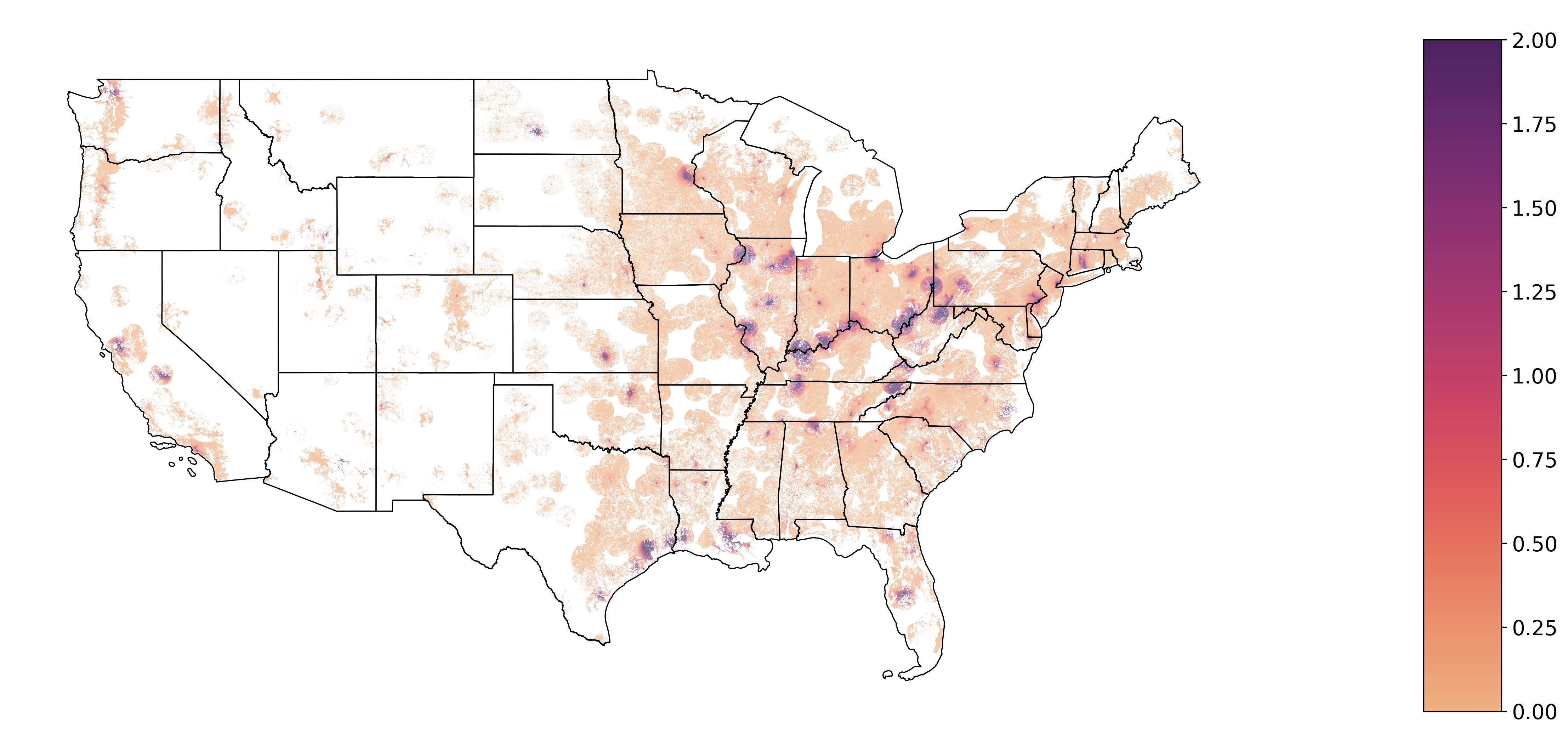}
        \caption{CEEM.}
        \label{fig:ceem_map}        
    \end{subfigure}
    \label{fig:CEEM}
    \caption{(a) Co-exposure of average wildfire smoke and total ground-level PM2.5 for September 2020 by AQI (Air Quality Index) threshold.
PM2.5 levels were hexified at level 8.
AQI thresholds for PM2.5 were used: Good = 50, Moderate = 100, Unhealthy = 200, and Very Unhealthy and Hazardous $>$ 200.
(b) The cumulative excess exposure across all air toxins (CEEM) for all Group 1 carcinogens in 2000.
CEEM is population masked, meaning unpopulated areas and areas with zero exposure are white.
Values over 1 indicate that exposure limits are exceeded at the mixture level even if individual carcinogens are below their limits.}
\end{figure}

Spatially indexing data to a common spatial extent allows for rapid and easy rendering of multidimensional maps.
Figure~\ref{fig:smoke_map} is a 2D map of co-exposure to PM2.5 air pollution and wildfire smoke in the US for September 2020.
Simultaneous mapping of exposures also allows for easy visualization of areas with differing exposure profiles.
In Figure~\ref{fig:smoke_map}, the areas with \textit{Good} air quality can be easily separated from the areas with \textit{Unhealthy} or \textit{Hazardous} air quality. It is also easier to identify areas that have air pollution exposures driven by wildfire smoke alone, non-smoke PM2.5, or a combination of smoke and non-smoke sources.
Figure~\ref{fig:ceem_map} shows the spatial distribution of CEEM for all carcinogens of Group~1 in the US in 2018.
Areas with a purple hue have a CEEM of~$>$1, indicating that exposure limits are exceeded at the mixture level, even if exposure limits for a single carcinogen are below the recommended limits.
In the western half of the US, there are areas with a CEEM of~$>$1 within most states.
A different pattern emerges around the Ohio River Valley, where high CEEM values tend to be located on or near state borders and are highly concentrated across the 981-mile stretch of the Ohio River.

\subsection{AAIRD for Unsupervised Clustering}

\subsubsection{Identifying Temporal Patterns of Elemental Carbon Exposure}
\begin{figure}[h]
    \centering
    \begin{subfigure}[b]{0.45\textwidth}  
        \centering
        \includegraphics[width=\linewidth]{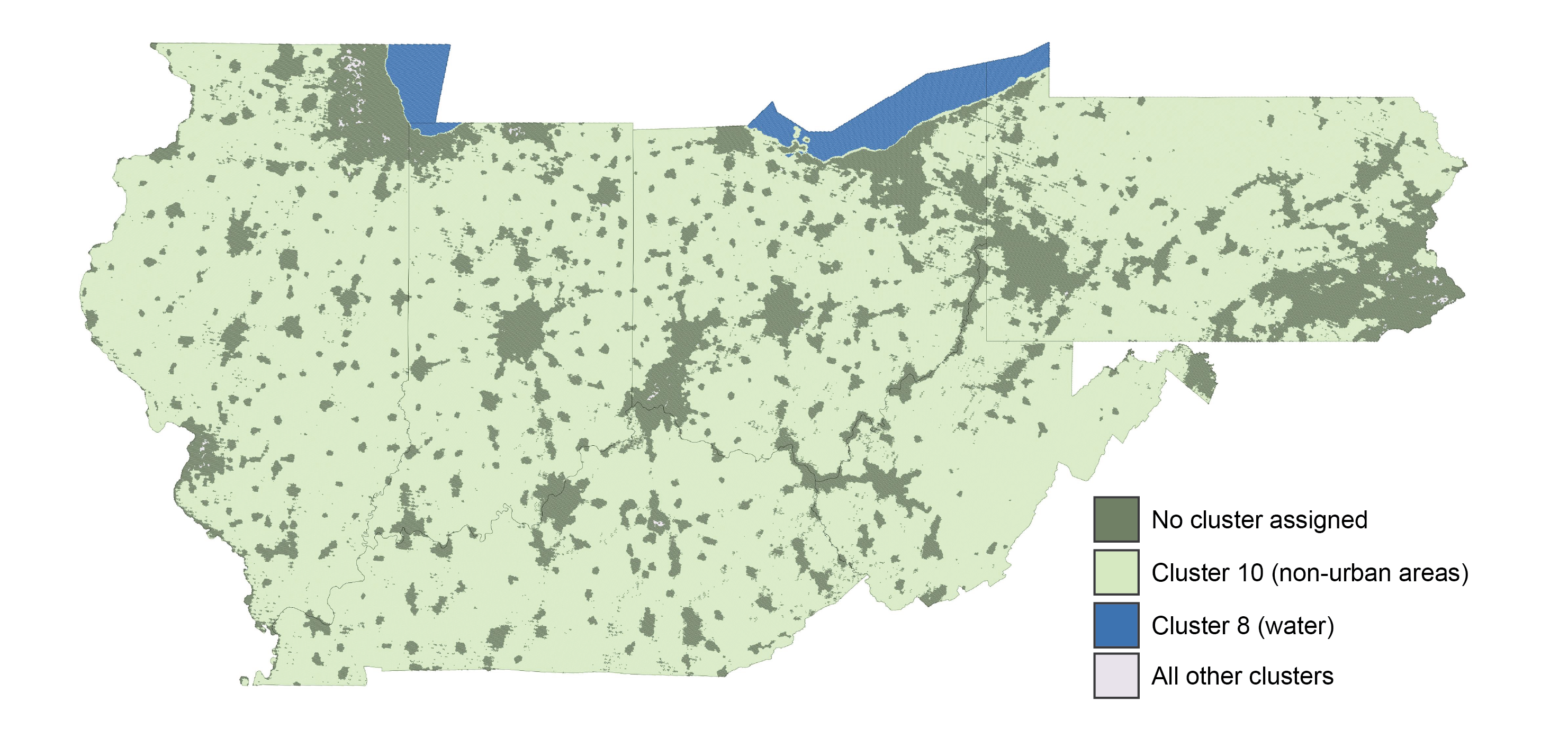} 
        \caption{Common clusters of elemental carbon exposure and noise.}
        \label{fig:CarbonMap}
    \end{subfigure}
    \hfill
    \begin{subfigure}[b]{0.45\textwidth} 
        \centering
        \includegraphics[width=\linewidth]{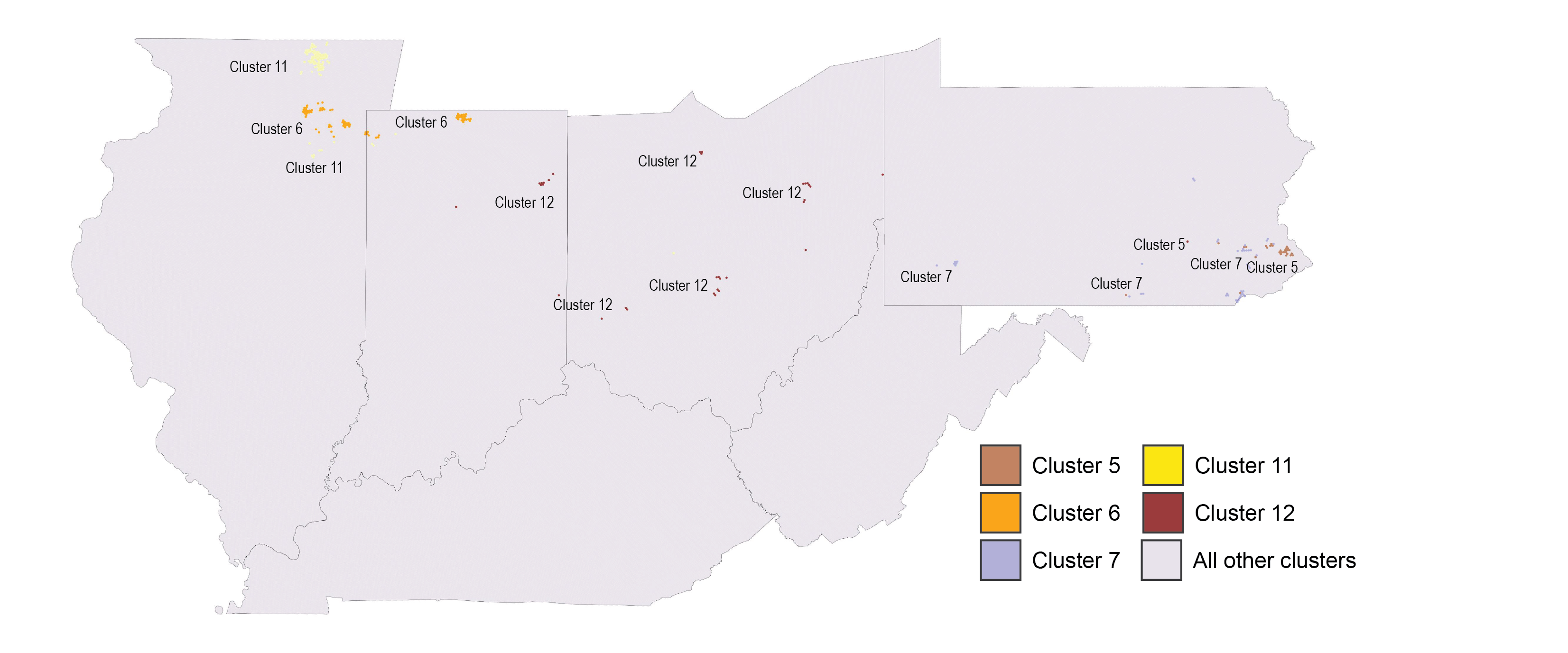} 
        \caption{Clusters of elemental carbon exposure that can be found across multiple locations.}
        \label{fig:ECExposure2}
    \end{subfigure}
    \hfill
    \begin{subfigure}[b]{0.45\textwidth}  
        \centering
        \includegraphics[width=\linewidth]{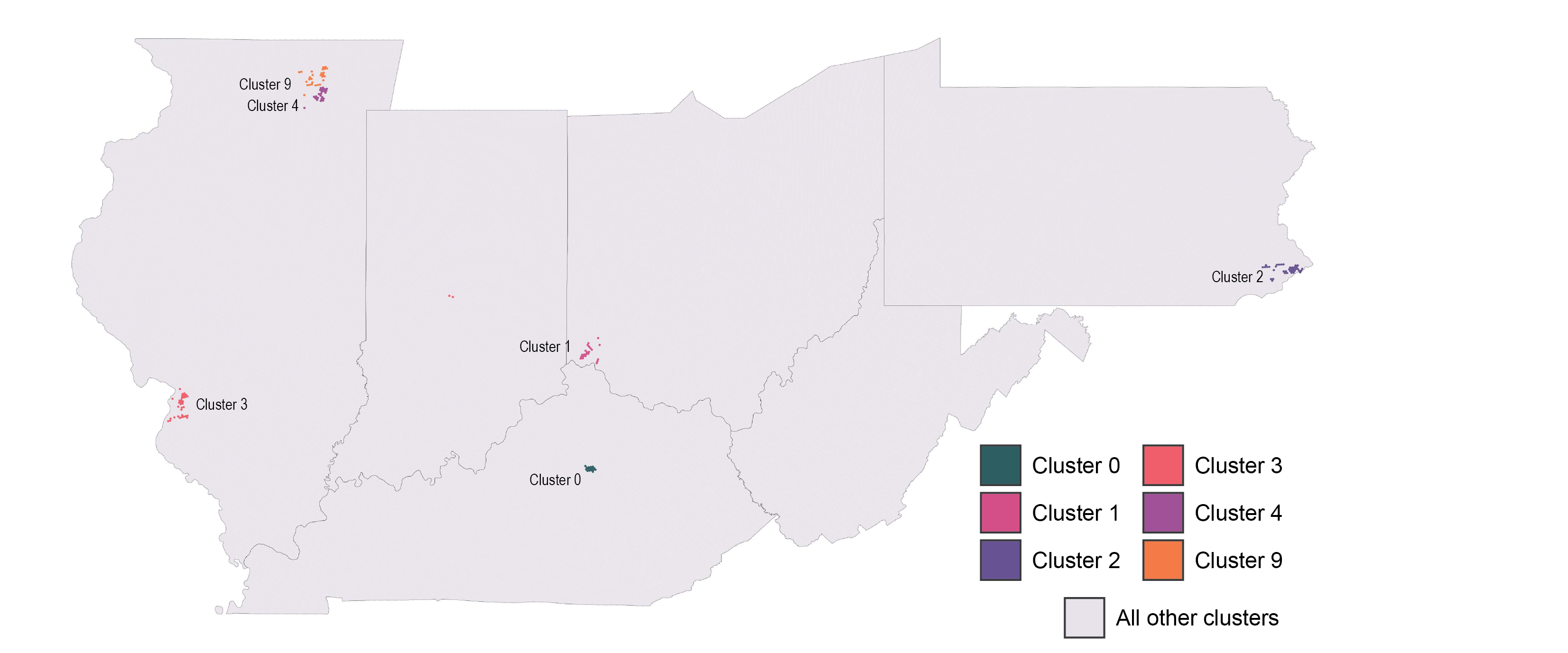} 
        \caption{Location-specific clusters of elemental carbon exposure.}
        \label{fig:ECExposure3}
    \end{subfigure}    
    \hfill
    \begin{subfigure}[t]{0.40\textwidth}  
        \centering
        \includegraphics[width=\linewidth]{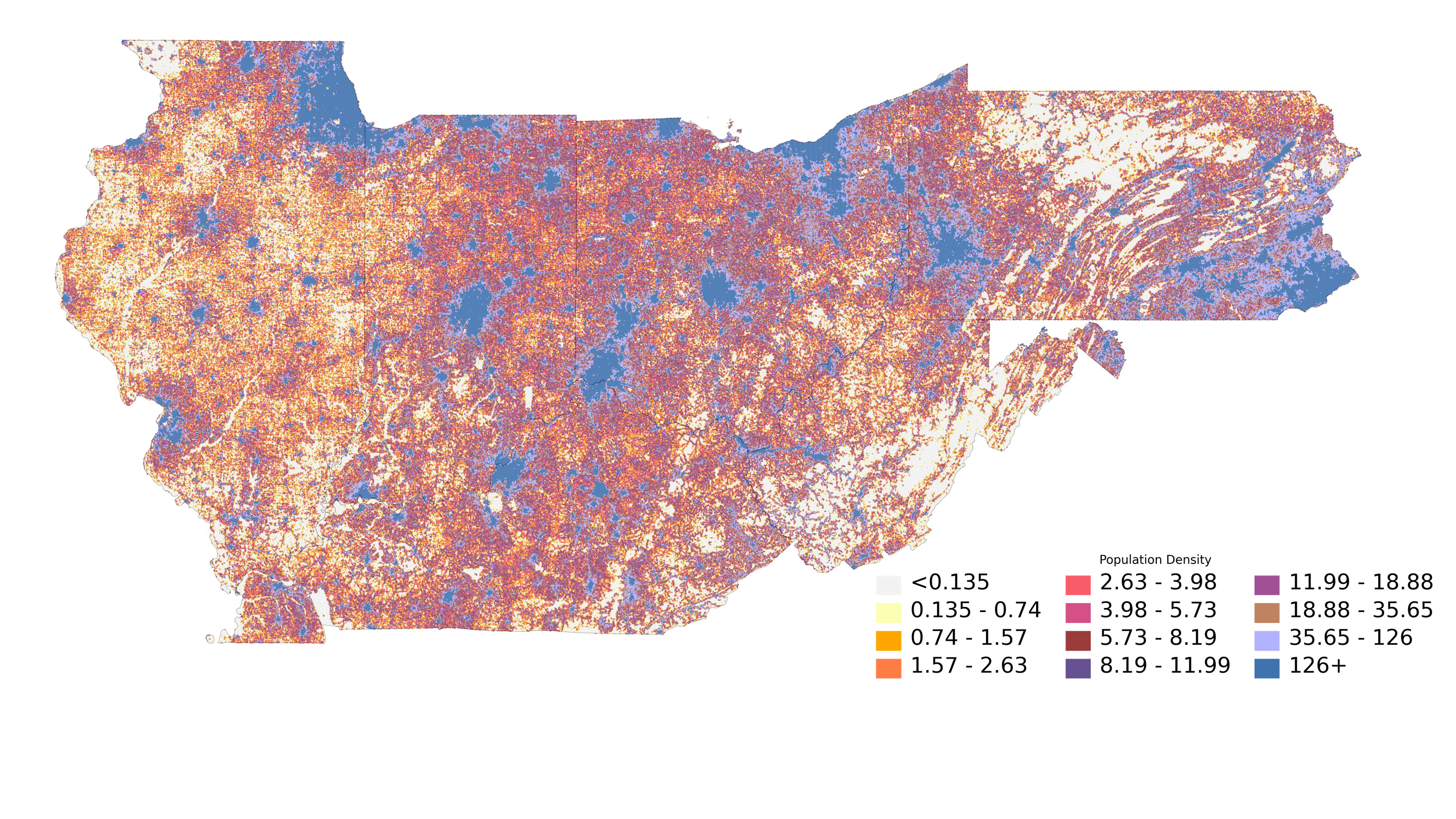} 
        \caption{Population density.}
        \label{fig:PopDensity}
    \end{subfigure}   
    \caption{HDBSCAN clustering results for elemental carbon from 2006 to 2019.
Two separate models are used to predict elemental carbon: one model for urban areas and one model for rural areas.
Clustering results show that patterns are driven by model differences.}
    \label{fig:mainCarbon}
\end{figure}

\begin{figure}[h]
    \centering
    \includegraphics[width=0.75\linewidth]{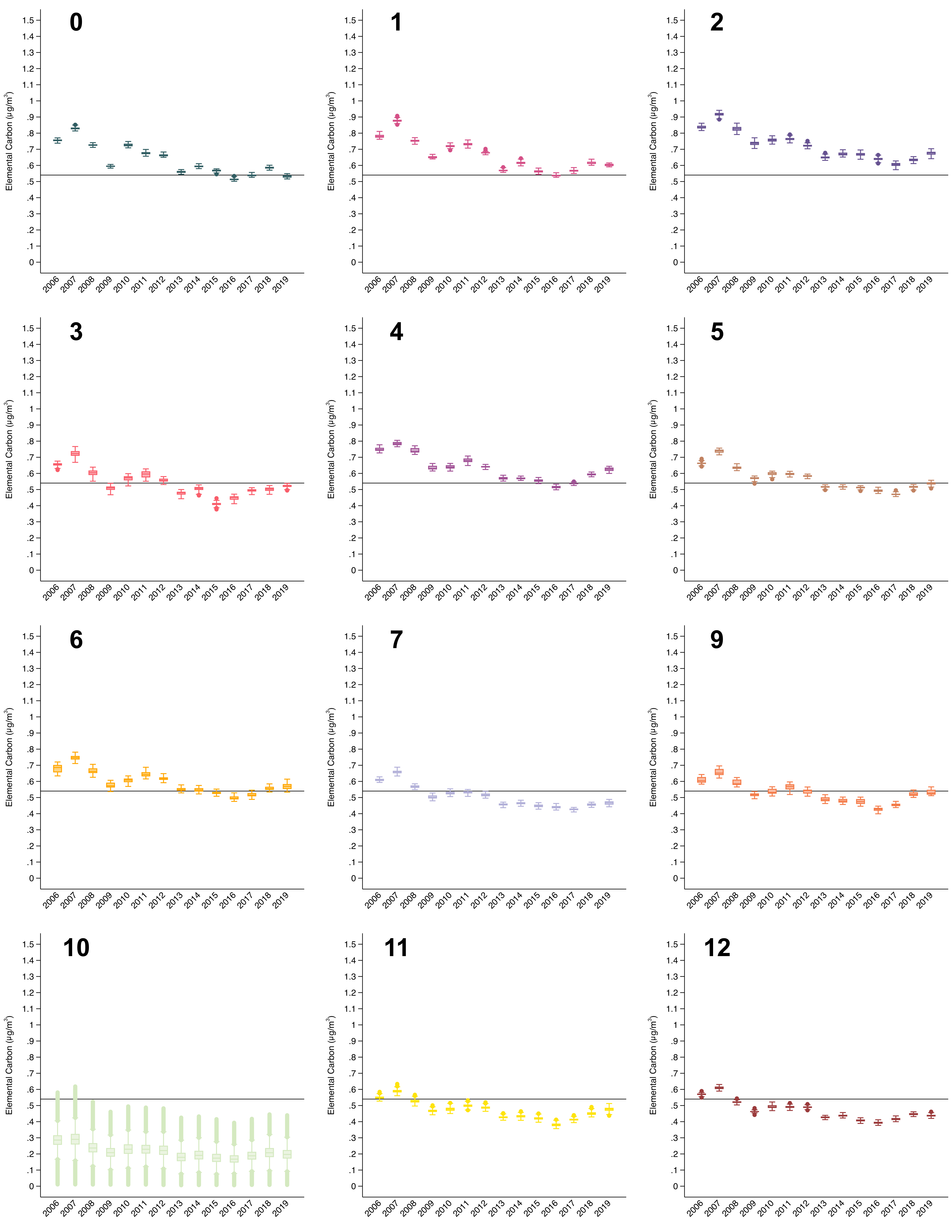}
    \caption{Box and whisker plots for elemental carbon levels over time and by cluster.
    The horizontal gray line represents the median concentration of elemental carbon in urban areas.}
    \label{fig:RR}
\end{figure}

HDBSCAN was used to perform aspatial clustering of temporal patterns in elemental carbon exposure (Figure \ref{fig:mainCarbon}).
The goal was to identify temporal patterns of exposure to elemental carbon between 2006 and 2019.
Thirteen distinct temporal patterns of elemental carbon exposure were identified, and hexes without clear patterns were classified as noise (see \textit{No cluster assigned} and \textit{Water} in Figure~\ref{fig:CarbonMap}).
Cluster 10 clearly covers most non-urban areas and captures hexes where concentration levels are consistently lower than the average for the region (Figure~\ref{fig:RR}).

An aspatial approach (i.e., spatial coordinates not considered) was used to enable the identification of similar temporal patterns of elemental carbon exposure in different locations.
Figure~\ref{fig:ECExposure2} highlights areas throughout the region with similar temporal patterns of exposure.
Although an aspatial approach was used, six clusters were localized to a single urban area (Figure~\ref{fig:ECExposure3}).
Four of these clusters (clusters 0, 1, 2, and 4) had the highest levels of elemental carbon concentration in the region between 2006 and 2012 (Figure~\ref{fig:RR}) and have consistently high values of elemental carbon compared to the other clusters.

Figure~\ref{fig:RR} shows the exposure of elemental carbon by group over time.
Most of the clusters show an overall decreasing trend in elemental carbon concentration, with spikes in exposure in 2007 and 2010.
Although they exhibit similar temporal trends, there are differences in the magnitude of exposure, the rate of decrease between 2007 and 2009, and the rate of increase between 2016 and 2018 by cluster.

\clearpage
\subsubsection{Identifying Patterns of Low-Dose Radiation Exposure}
\begin{figure}
    \centering
    \begin{minipage}[b]{0.45\textwidth}
        \begin{subfigure}[t]{\textwidth}
            \centering
            \includegraphics[width=\linewidth]{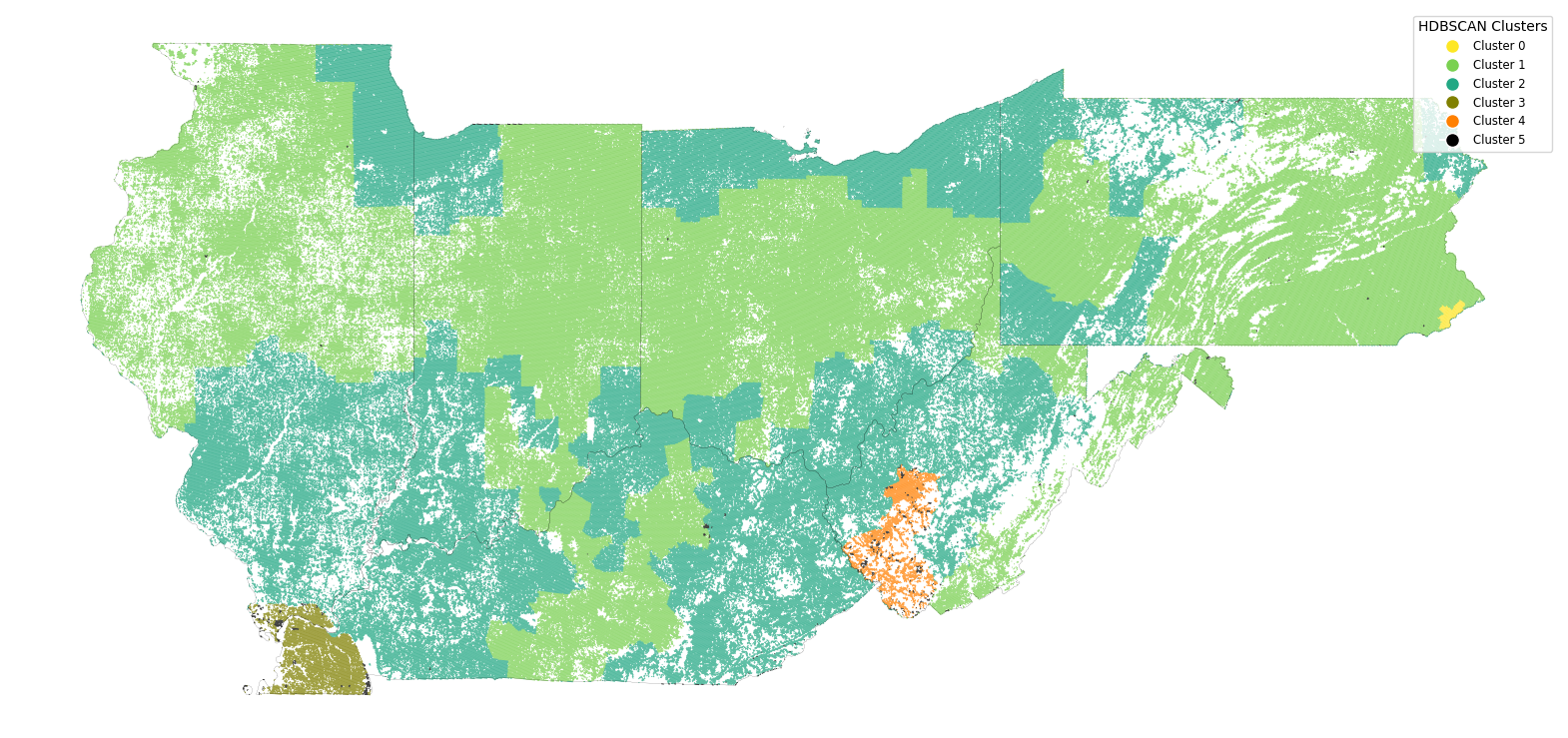} 
            \caption{HDBSCAN clustering results for radiometric data.}
            \label{fig:Solar}
        \end{subfigure}
        
        \vspace{1em}
        
        \begin{subfigure}[b]{\textwidth}
            \centering
            \includegraphics[width=\linewidth]{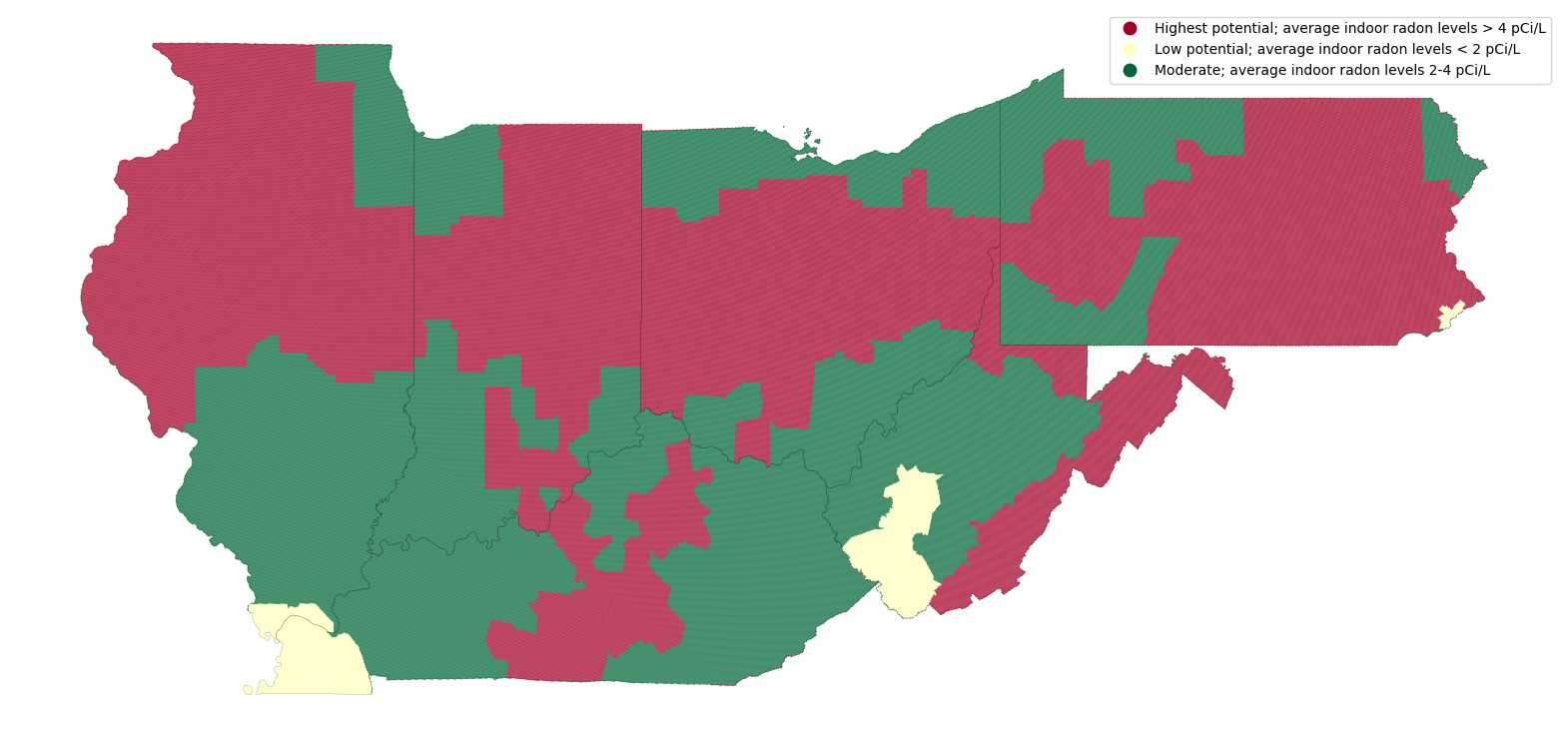} 
            \caption{EPA indoor radon zones.}
            \label{fig:Radon}
        \end{subfigure}
    \end{minipage}
    \hfill
    \begin{minipage}[b]{0.45\textwidth}
        \begin{subfigure}[b]{\textwidth}
            \centering
            \includegraphics[width=\linewidth,height=\textheight,keepaspectratio]{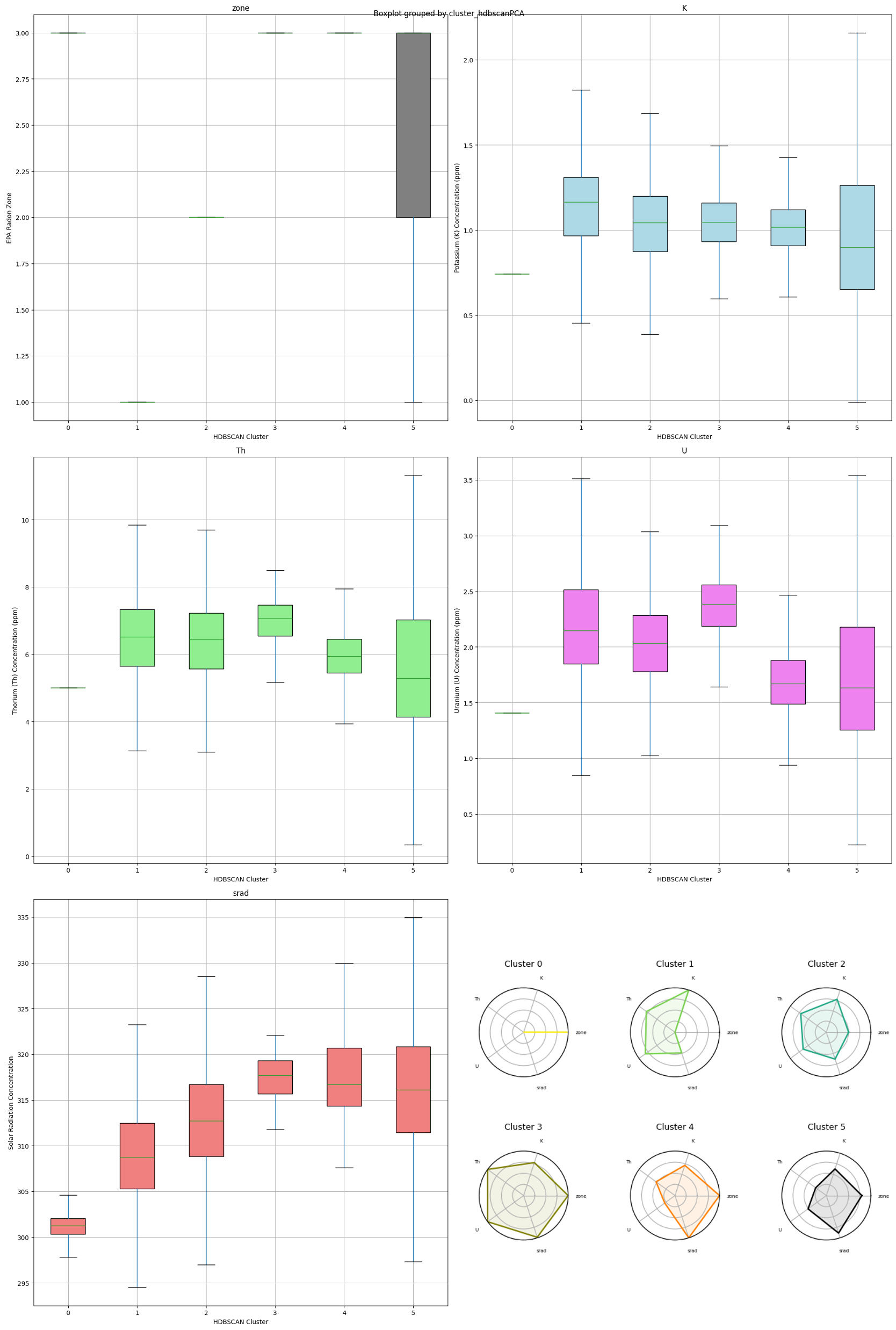} 
            \caption{Boxplot based on hexified clusters.}
            \label{fig:BoxRadar}
        \end{subfigure}
    \end{minipage}
    \caption{Radiometric clustering results for the Ohio river valley in 2018.
(a) Spatial distribution of cluster assignment.
(b) County-level EPA indoor radon zones.
(c) Boxplot of hexified clusters.}
    \label{fig:rad-cluster-plots}
\end{figure}

Figure~\ref{fig:Solar} shows the clustering results for radiometric data.
Unlike the elemental carbon example, spatial coordinates were included, and all measures were from 2018.
As in the other examples, the clusters are parsed into noise and meaningful signals.
Although data were indexed to hex8, EPA indoor radon estimates were not spatially downscaled, and all hexes within a county had the same value for potential elevated indoor radon (Figure \ref{fig:Radon}).
The clusters are largely influenced by the indoor radon potential, which was coded following the EPA scheme, where zone 1 has the highest potential and zone 3 has the lowest potential.
Figure~\ref{fig:BoxRadar} shows that the hexes classified as clusters 0, 3, and 4 are found in areas with low potential for indoor exposure to radon.

Compared to the other clusters, cluster 0 has the lowest levels across all radiometric measures.
Cluster 3 has a higher thorium and uranium content in the soil than cluster 4, likely indicating that, although both areas have a low potential for indoor radon exposure, there are differences.
Cluster 5 is the noise cluster and, in this example, it can be interpreted as anomalous areas where hexes with these classifications are sprinkled throughout the Ohio River Valley but are mainly concentrated in counties classified as EPA Zone 1 areas.
This suggests that the EPA classification in these areas may underestimate potential radon exposure.

\subsubsection{Identifying Patterns of Exposure Across Multiple Exposure Domains: Characterizing the Exposome}
\begin{figure}[h]
    \centering
    \begin{subfigure}[b]{0.49\textwidth}
        \centering
        \includegraphics[width=\linewidth]{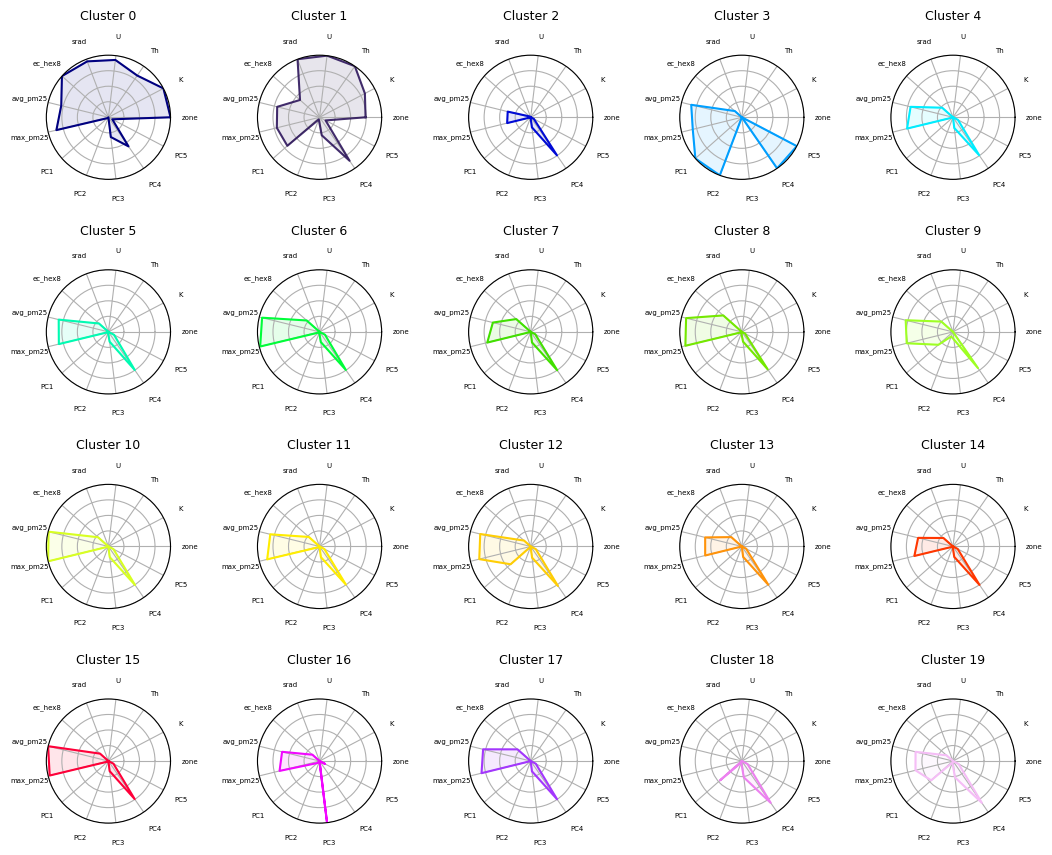}
        \caption{Air pollution data, radiometric data, and RSEI PCs 1--5.}
        \label{fig:Radar1}
    \end{subfigure}
    \hfill
    \vline
    \begin{subfigure}[b]{0.49\textwidth}
        \centering
        \includegraphics[width=\linewidth]{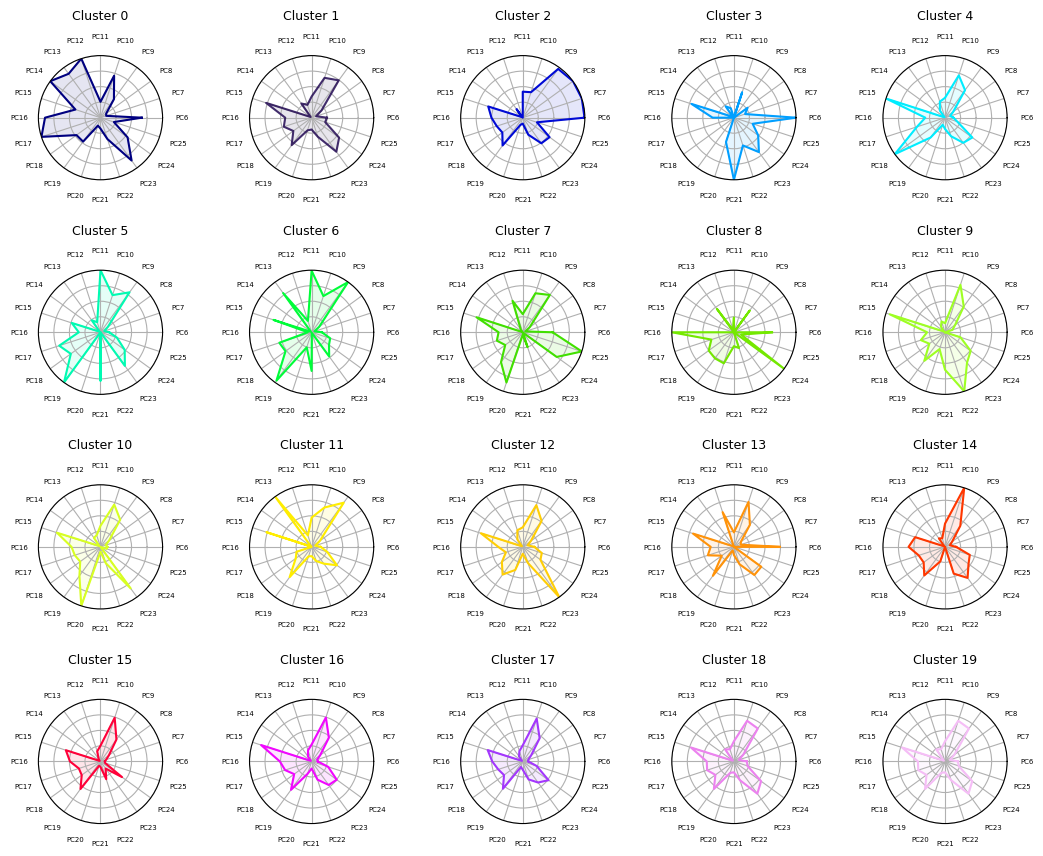}
        \caption{Exposomic signatures for RSEI PCs 6--25.}
        \label{fig:Radar2}
    \end{subfigure}
\caption{Exposure signatures by cluster.}
\label{fig:Radar}
\end{figure}

The HDBSCAN clustering identified 19 distinct patterns of co-exposure in the Ohio River Valley in 2018.
Radar plots were used to visualize the unique exposomic signature of each group (Figure~\ref{fig:Radar}).
To facilitate interpretation, the plots were split into exposure groups: (a) air pollution, radiometric, and RSEI PCs 1--5 (Figure~\ref{fig:Radar1}) and (b) RSEI PCs 6--25 (Figure~\ref{fig:Radar2}).
RSEI PCs 1--5 explain 39.2\% of the total variance in chemical exposure, and PCs 6--25 explain the additional 29.2\%.

\begin{figure}[h]
    \centering
    \begin{subfigure}[b]{0.45\textwidth}
    \centering
        \includegraphics[width=\linewidth]{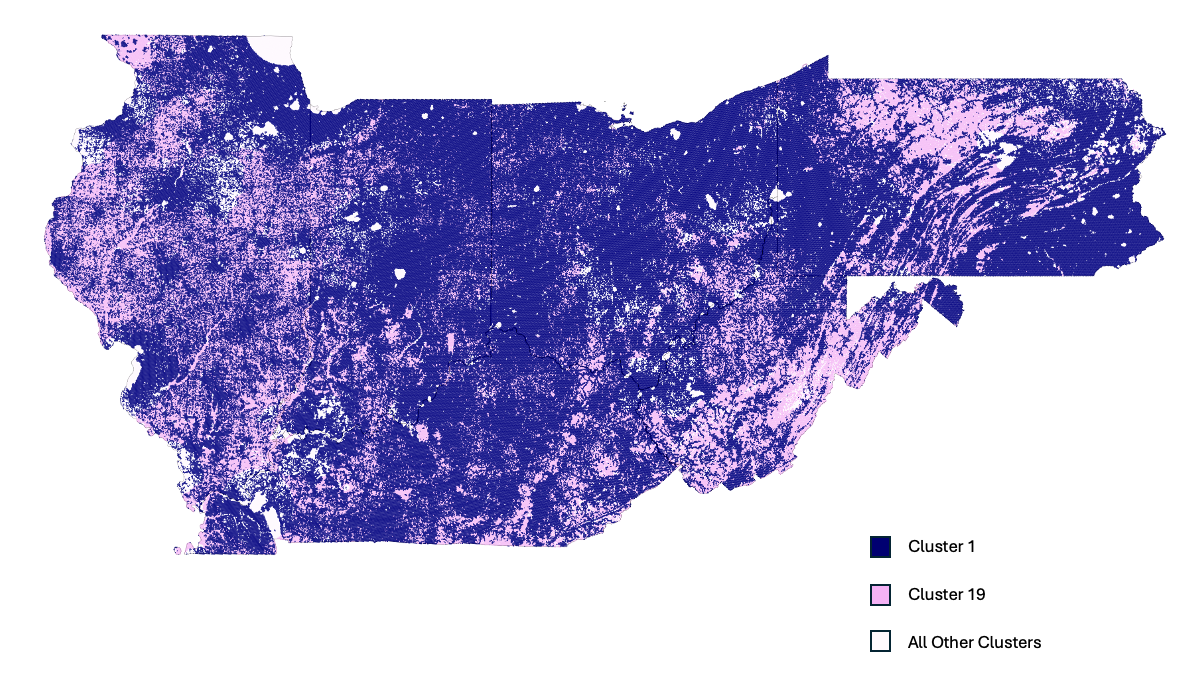}
        \caption{Clusters 1 and 19.}
        \label{fig:Clust1_19}        
    \end{subfigure}
    \hfill
    \begin{subfigure}[b]{0.45\textwidth}
        \centering
        \includegraphics[width=\linewidth]{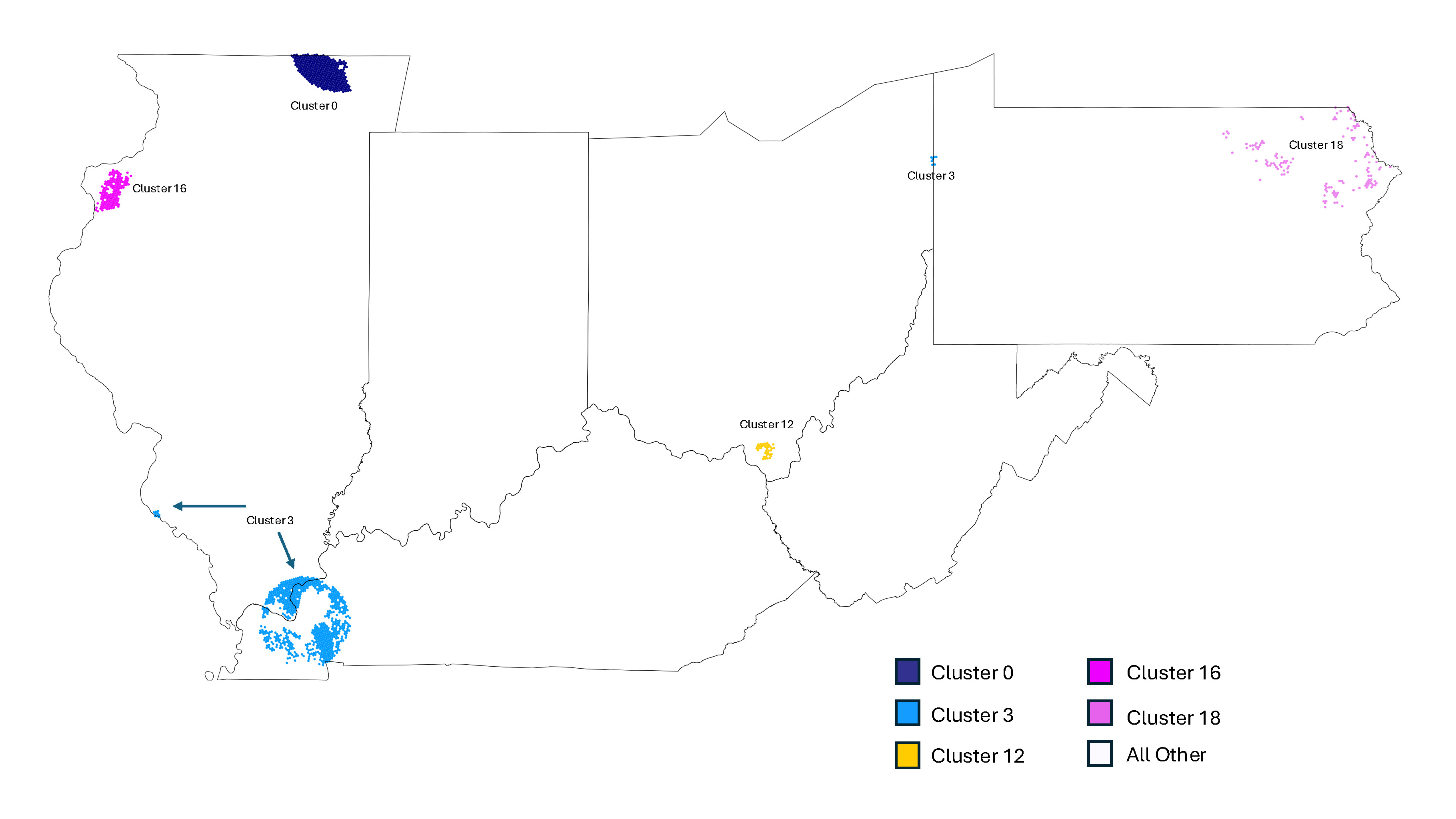}
        \caption{Clusters 0, 3, 12, 16, and 18.}
        \label{fig:Clust0_3_12_16_18}        
    \end{subfigure}    
    \hfill
    \begin{subfigure}[b]{0.45\textwidth}
        \centering
        \includegraphics[width=\linewidth]{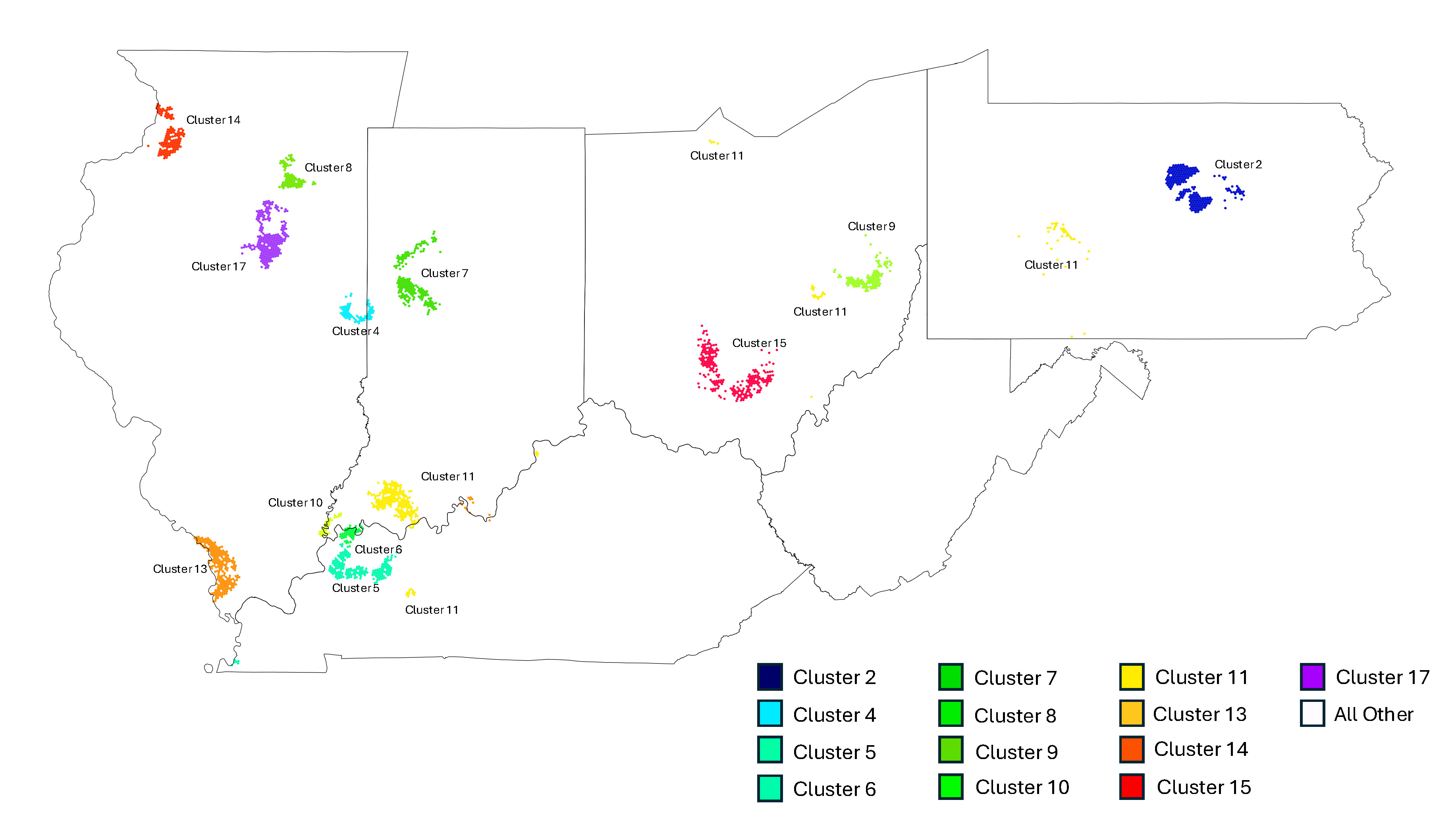}
        \caption{Clusters with distinct signatures of chemical exposure (PC4) and PM2.5.}
        \label{fig:ClustPM25_PC4}        
    \end{subfigure}   
\caption{Exposomic clusters.}
\label{fig:ExpoClusters}
\end{figure}

Three distinct patterns across all 19 clusters become evident when assessing the relative magnitude of exposure across the first set of variables.
First, clusters 1 and 19 hexes in unpopulated areas were largely classified as cluster 19 or noise (Figure~\ref{fig:Clust1_19}).
Clusters~1 and 19 were the largest in terms of number of hexes, with 660,971 hexes classified as cluster~1 ($\sim$467,306~km$^2$) and 183,304 classified as cluster~19 ($\sim$129,595~km$^2$) (Figure~\ref{fig:Clust1_19}).
A total of 58,379 hexes were classified as noise (35,195~km$^2$).
Outside of these classifications, cluster 3 is the largest and covers 1,936~km$^2$ of land in Kentucky, Illinois, and Pennsylvania.
Second, the clusters mapped in Figures~\ref{fig:Clust0_3_12_16_18} and \ref{fig:ClustPM25_PC4} can be differentiated by the patterns in the average PM2.5, maximum PM2.5, and RSEI PC~4.

Clusters 0, 3, 12, 16, and 18 have distinct patterns of radiometric exposure and RSEI exposure, whereas clusters 2, 4, 5--11, 13--15, and 17 have similar exposure signatures that vary only by the magnitude of PM2.5 and RSEI PC~4.
The loadings for RSEI PCs 6--25 further differentiate the clusters, and most clusters have unique signatures for these variables.
A table of the most important chemical per PC can be found in \hyperref[sec:supplement-3]{Supplement 3}.
Cluster 6 has high PC~11 and PC~19 loading, both of which have most of the important features for IARC Group 1 (known carcinogen) or IARC Group 2a (probable carcinogen).
The cluster covers a portion of rural Henderson County, which had an age-adjusted cancer incidence rate of 417.3 per 100,000 (95\% CI 384.4--452.5) from 2019 to 2023 \cite{StateCancerProfiles2025}.
This is similar to the age-adjusted incidence rate in the US during the same period of time.

\subsection{AAIRD for Linking Exposomic Measures to Residential Histories for Cancer Patients}
We created a pipeline for creating point-level linkages of residential history data, which is provided by the SEER program, to air pollution exposure data.
These links will enable future researchers to assess the relationship between longitudinal exposure measures and cancer incidence, response to treatment, and survival.
The exposome linkage pipeline was created using data from the Louisiana, New Jersey, Kentucky, and Iowa cancer registries, which had previously been linked to LexisNexis residential history data \cite{tatalovich2022assessment}.

\begin{figure}[h]
    \centering
    \begin{subfigure}[b]{0.49\textwidth}
        \includegraphics[width=\linewidth]{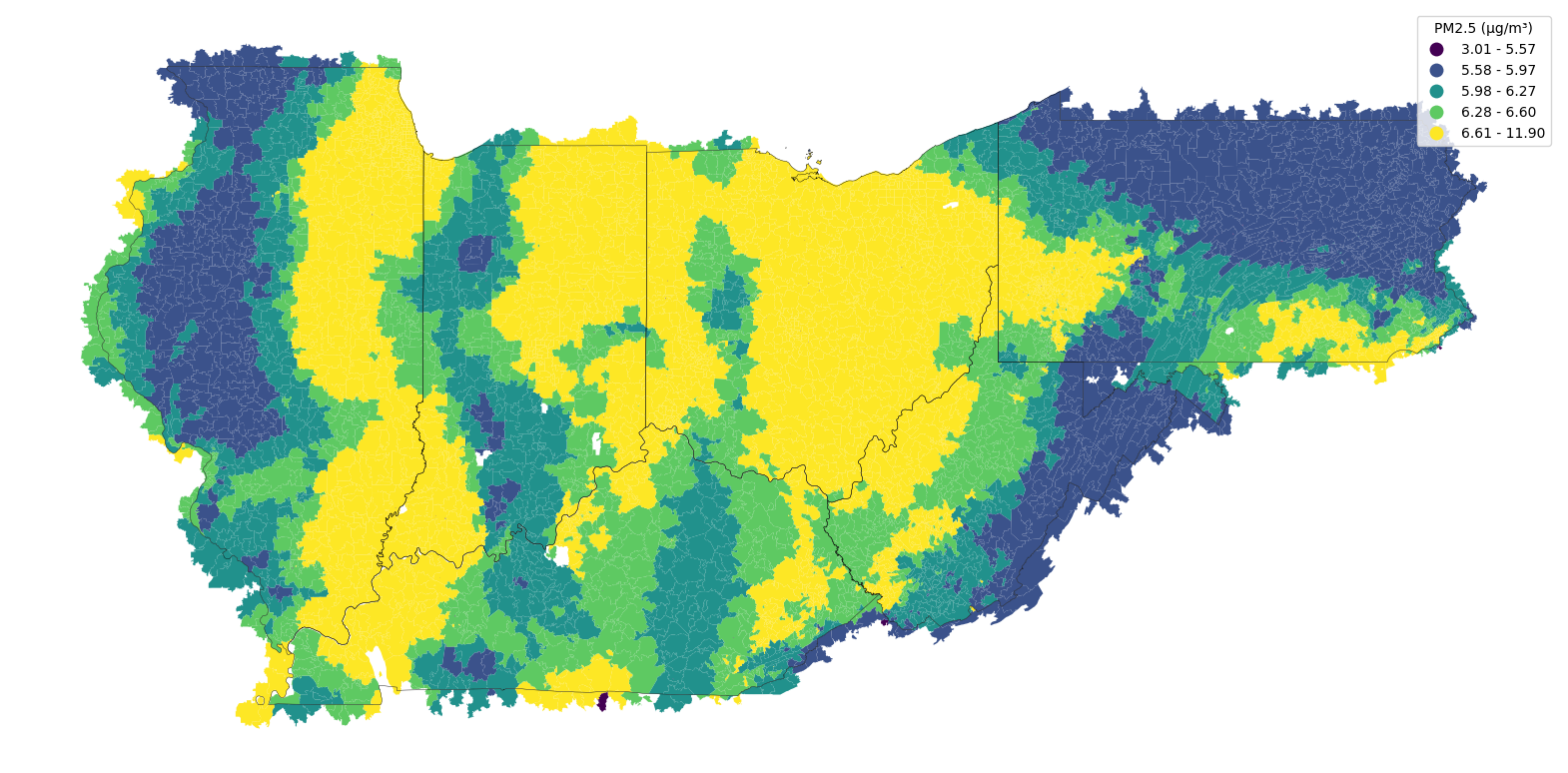}
        \caption{Quintiles of average daily PM2.5 exposure in 2018 estimated at the hex8 level and aggregated by zip code to facilitate linkages to addresses geocoded to a zip centroid.}
    \label{fig:ZipAvg}      
    \end{subfigure}
    \hfill
    \centering
    \begin{subfigure}[b]{0.49\textwidth}
        \includegraphics[width=\linewidth]{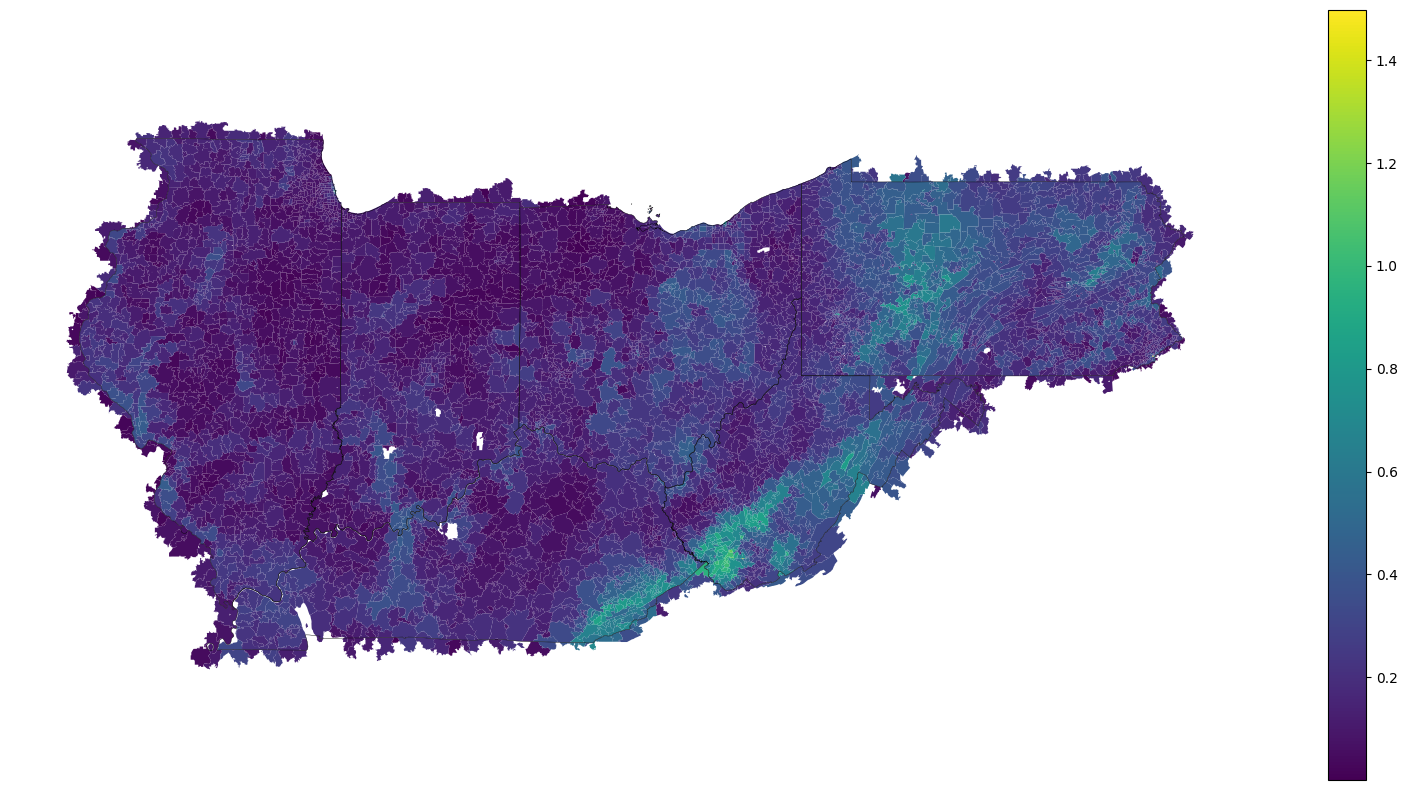}
        \caption{Standard deviation of average daily PM2.5 exposure in 2018 estimated at the hex8 level and aggregated to the zip code to facilitate linkages to addresses geocoded to a zip centroid.}
    \label{fig:ZipStd}      
    \end{subfigure}    
\label{fig:Zip}
\caption{Ground-level, real-time modeling of PM2.5 pollution with Senseiver and aggregated to the zip code level.}
\end{figure}

Address data from 1995 to 2024 were geocoded using the ADDRESS (Automated Determination of Detailed Regional and Exact Spatial Segments) geocoder developed at Oak Ridge National Laboratory.
All addresses were geocoded to a point location and to the H8 level.
This simple spatial indexing allowed for easy mapping between residential address and exposomic measures.
Residential history data were available for a total of 927,507 cancer patients from the states mentioned above with 2,982,450 addresses in the US between 1995 and 2024.
Of these addresses, 1,871,626 were geocoded to the point level, and 1,060,257 were geocoded to the zip code.
The mapping from hex to zip code is quite simple and enables exposure measures at different spatial resolutions to be assigned based on the geocode level (Figure~\ref{fig:ZipAvg}).
A full description of the linkage protocol can be found in prior work.
Although a complete analysis is beyond the scope of this article, researchers can apply for access to CONNECT data through the SEER registries.

\section{Discussion}
The heterogeneity of external exposomic data presents one of the most significant barriers to advancing environmental epidemiology and exposomic research \cite{zheng2020design, hu2023methodological}.
Inconsistencies in data quality and format (e.g., satellite-derived estimates versus ground-based sensors), metadata structures, and ontologies hinder interoperability and large-scale integration.
These inconsistencies often require researchers to spend substantial effort cleaning, standardizing, and reconciling the data before meaningful analysis can occur.
This work outlines C-HER and describes how our AAIRD ecosystem can be used to overcome these challenges and produce data that can be used for exposure modeling, thematic mapping, unsupervised clustering, and exposomic analyses.
This is especially attractive when health impact assessments and interventions are needed in near real time.

AAIRD supports data-driven decision-making by providing timely and quantifiable information on environmental risks and can be used to create targeted health interventions and policy actions \cite{wu2025small}.
Combined with the Senseiver architecture and built-in uncertainty quantification at inference time, the C-HER ecosystem provides a framework for near real-time exposure modeling, which is a critical capability for protecting public health.
In contrast to other popular models used for exposomic and environmental epidemiology research \cite{di2019ensemble}, our method uses a few important features to make fine-grained, data-adaptive predictions from data across the US.
Although we demonstrated feasibility using a single model, C-HER's AAIRD framework is compatible with any statistical or deep learning model.

Harmonization frameworks such as C-HER offer a promising solution to improving reproducibility, increasing the rate of scientific discovery, and providing data-driven insights into the complex relationships between the environment and health \cite{kush2020fair, stingone2025unlocking}.
C-HER also offers a solution to privacy-related issues by enabling distributed data access while preserving data sovereignty without requiring physical centralization.
The work described here is an example of how the FAIR C-HER AAIRD can be shared with partners such as the National Cancer Institute.
Although not mentioned here, a similar process is also being implemented for the US Department of Veterans Affairs.
Using the same datasets across institutions improves trust in exposomic research findings by allowing for independent replication and consistency in results across diverse populations.
Sharing data through the C-HER ecosystem also promotes transparency through common provenance tracking and standardized quality controls to enable open methodological comparisons across studies.
Ultimately, this will strengthen public and scientific confidence in exposomic findings and accelerate translation into actionable public health insights.

Although solutions such as C-HER overcome barriers in accessing a large amount of exposomic data, they do not solve issues related to interpretation of high-dimensional signatures of exposure.
We provide a simplistic representation of exposomic signatures by clustering data for air pollution, radiometric, and toxic releases from across the Ohio River Valley.
Although these signatures may be related to health outcomes, they are difficult to interpret and link to underlying biologic mechanisms.
Additionally, they may capture artifacts in data measurement or scaling, making it unclear whether the observed correlations are real or the result of some unrelated data process.
Clustering results can also be sensitive to algorithm choice, distance metrics, and parameter settings.
The benefit of C-HER is that broad sharing of standardized and harmonized data will facilitate more thorough analyses by allowing independent researchers to test and validate findings across multiple datasets,
ultimately strengthening our understanding of the links between exposure mixtures and health outcomes.

\newpage

\bibliography{references}  





\newpage

\appendix

\renewcommand{\figurename}{Supplementary Figure}
\renewcommand{\thepage}{\Alph{section}-\arabic{page}}
\setcounter{page}{1}
\setcounter{figure}{0}

\section{Supplement 1. Challenges and Solutions for Converting Data to Hex8}
\label{sec:supplement-1}

The specific process to convert arbitrary data sources to H3 depends on various factors.
Input data may be raster or vector in various file formats and resolutions using different projections.
For this reason, workflows must consider different data semantics.
For example, counts, averages, and per-capita values are structurally similar but require different H3 conversion strategies.
Data size varies from small, single-band rasters to huge datasets with high spatial and temporal resolution and extent.
For such large datasets, a direct solution often requires more memory than available and can result in long execution times for codes that fail due to memory overrun after executing for hours.
Some of the useful libraries offer only partial solutions to our conversion problem, so we must augment or cobble together multiple, partial solutions---an approach that can place data integrity at risk.

When converting a new dataset, the first step is to examine the data to understand its structure.
Tools such as \texttt{ls} and \texttt{du} provide an initial view of data size for individual files and collections.
Text formats such as \texttt{csv}, \texttt{head}, and \texttt{wc} let us see the structure and object count.
For binary formats, we rely on special-purpose tools such as \texttt{gdalinfo} and \texttt{parquet-tools} to expose that information.
Once we understand the size, structure, and meaning of the input data, we prepare the data by applying operations to standardize it.
This may include re-projecting, converting rasters to \texttt{tiff} format, discarding data outside of desired extents, and generally dealing with data idiosyncrasies.
For raster datasets, \texttt{gdal-warp} and \texttt{gdal-translate} may be sufficient, but preparation for other datasets will require customized Python solutions that leverage libraries such as GeoPandas and Rasterio.

\begin{figure}[h]
    \centering
    \includegraphics[width=0.5\linewidth]{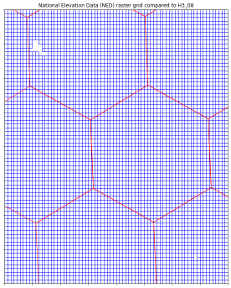}
    \caption{NED 30 resolution compared with H3, level 8.}
    \label{fig:SFig2}
\end{figure}

We have explored several strategies for converting the prepared data to H3.
The first strategy uses the \texttt{geo\_to\_h3\_aggregate()} method from the H3-Pandas library.
This method operates on a GeoDataFrame object that contains the input data and takes the resolution and aggregation operation as arguments.
An H3 index is added to the dataframe by assigning each row of the input to an H3 cell of the specified resolution.
Rows are then grouped by H3 index, and the aggregation operation is applied to each group.
The \texttt{geo\_to\_h3\_aggregate()} method abstracts the input geometries to a single point, called the \textit{centroid}, and is not appropriate for all datasets.
We have used \texttt{geo\_to\_h3\_aggregate()} primarily for situations in which the input geometry has significantly higher resolution than that of the target H3 hexagons.
The National Elevation Dataset (NED), with an approximately 1 arc-second resolution (roughly 30~$\times$~30~m), is a good example.
Supplementary Figure~\ref{fig:SFig2} shows the relative size of the NED 30 raster pixels (shown in blue) compared with level-8 H3 hexagons (shown in red).

Another strategy is to use the \texttt{h3\_polyfill()} method from the H3-Pandas library.
The \texttt{h3\_polyfill()} method works by applying a value from an input geometry to any H3 cell with a centroid contained in that geometry.
This technique is appropriate for situations in which input data are more coarse than the target H3 resolution.
One example of data that fits this case comes from the ERA5 dataset.
ERA5 provides re-analysis of climate and weather data for the past 8 decades.
We focus on hourly temporal resolution and the Mean Sea Level Pressure variable, which is available at 0.25\degree~$\times$~0.25\degree spatial resolution.
A comparison of this spatial resolution versus level-8 H3 is shown in Supplementary Figure~\ref{fig:SFig3}.
The ERA5 raster pixels are shown in blue, and the level-8 H3 are shown in red.

\begin{figure}
    \centering
    \includegraphics[width=0.5\linewidth]{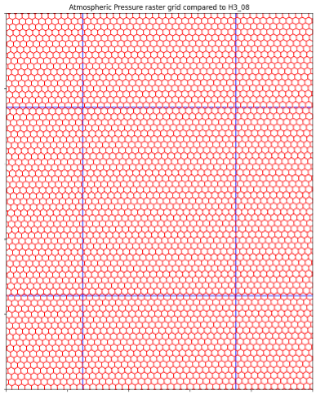}
    \caption{Spatial resolution for ERA5 mean sea level pressure compared to H3, level 8.}
    \label{fig:SFig3}
\end{figure}

The above H3-Pandas methods are suitable for cases in which there are consistent size disparities between input geometry and it is reasonable to treat the smaller objects as points without introducing significant error.
However, for addressing situations in which objects in the dataset have sizes similar to the target H3 hexagons, or in which the dataset has both large and small objects (e.g., census tract or block group, where object size depends on population), the H3-Pandas methods are insufficient.
The Daymet data are an example of this case.
Daymet provides a daily modeled snapshot of seven weather variables covering North America at a 1~$\times$~1~km resolution.
As shown in Supplementary Figure~\ref{fig:SFig4}, the pixels in the Daymet raster (blue) cover a similar area as the level-8 H3 hexagons (red).

\begin{figure}
    \centering
    \includegraphics[width=0.5\linewidth]{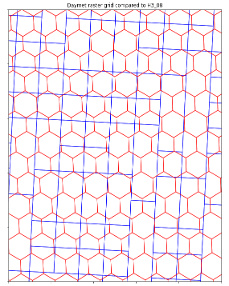}
    \caption{Relative spatial resolution of the Daymet raster grid compared to H3, level 8}
    \label{fig:SFig4}
\end{figure}

One way to address these situations is to use the \texttt{area\_interpolation()} method from the Pysal Tobler Python library to compute spatially weighted values for each hexagon using a weighted average of complete and partial pixels that overlap that hexagon.
We found this approach challenging in several ways.
First, the interface obscures some of the details about how the interpolation is accomplished.
It offers intensive, extensive, and categorical options for interpolating, but the results from specifying intensive versus extensive variables were not always as expected.
The structure that resulted from categorical values provided more detail than was needed and required additional processing to resolve.

Finally, the \texttt{area\_interpolation()} interface did not offer any obvious way to control how source data was allocated to hexagons when chunking the processing (discussed below).

Finally, we explored another interpolation strategy with the GeoDataFrame \texttt{overlay()} method from the GeoPandas library.
This technique avoids some of the challenges that we encountered with the Tobler library.
For this method, we generate geometries for the target hexagons and use \texttt{overlay()} to obtain a frame with all of the fragments created from intersecting the two sets of geometries as well as the relative area of each fragment with respect to both the original geometry and the target hexagon.
This provides us with a map that can be reused for all input data with the same geometry.
For each such input frame, we join the map and then group by hexagon to generate the set of fragments that makes up each hexagon.
We then apply an appropriate aggregation operation to those sets to produce the converted H3 data.

For larger datasets, we have encountered two limitations that required a change to the basic approach outlined here.
The first involves inadequate system memory, which arises from attempting to process too much data in memory at one time.
To account for this, we have used a tiling or chunking approach to process partial data in a streaming fashion.
Here, we split the spatial data into regular chunks, read one chunk at a time, process that chunk, and then free the associated memory before proceeding to the next chunk.
We found that it was best to divide the space according to the output hexagons and to ensure that all related input data was present while processing each chunk, which required some overlap between input chunks but prevented inaccuracies along the edges of the tiles.

For the largest datasets, typically involving higher temporal resolution like daily or hourly, we found that even when managing memory appropriately, the run time of the conversion process could still be quite long. Leveraging a computing cluster was necessary for parallelizing the work and reducing the run time.

\newpage

\section{Supplement 2. Data Management Plan for the Centralized Health and Exposomic Resource Ecosystem}
\subsection{Introduction} This data management plan outlines the strategies for managing data within the Centralized Health and Exposomic Resource (C-HER) ecosystem to ensure efficient data handling, security, and compliance with relevant standards.
The C-HER ecosystem utilizes open-source software such as MinIO, PostgreSQL, Prefect, and Python.
The use of open-source tools and sharing of these code bases (data ingestion code) will enable validating results and preserving any record of transformation that was performed on the data.
Optimizing the reuse of data, the C-HER ecosystem adheres to principles of findability, accessibility, interoperability, and reusability (FAIR).
Datasets are distributed under a Creative Commons License as appropriate for each dataset.

\subsection{Data Description} 
\subsubsection{Data Types} 
\begin{itemize} 
    \item \textbf{Raster Data}: Includes satellite images, aerial photographs, and derived raster products.
    \item \textbf{Vector Data}: Geographic features such as boundaries, roads, and points of interest.
    \item \textbf{Tabular Data}: Data stored in structured formats within PostgreSQL, including metadata, user information, and analytical results, including H3 hexes.
    Tabular data should not contain geometry but be indexed to the respective geometry in the boundaries tables.
    \item \textbf{Large Source Files}: Raw data files stored in MinIO, including large datasets used for processing and analysis.
    \item \textbf{Data Ingestion Code}: The code used to ingest the data will also be stored in the database to enable quick determination of how data was inserted into the CHER ecosystem.
\end{itemize} 
    
\subsubsection{Data Format and Scale} 
\begin{itemize} 
    \item Raster and vector data are stored in formats such as GeoTIFF, Shapefile, and GeoJSON.
    \item Tabular data are managed in SQL format within PostgreSQL databases.
    \item Large source files are maintained in their native formats to preserve integrity and facilitate access.
\end{itemize} 
    
\subsection{Roles and Responsibilities} 
\begin{itemize} 
    \item \textbf{Data Managers}: Oversee data storage architectures and data life cycle management.
    \item \textbf{System Administrators}: Maintain the IT infrastructure and ensure data security.
    \item \textbf{Researchers and Analysts}: Generate, process, and analyze data while adhering to standard protocols.
    \item \textbf{Standards Committee}: Reviews and edits the standards document at scheduled intervals (e.g., twice per year).
\end{itemize} 

\subsection{Data Storage and Preservation} 
\subsubsection{Storage Solutions} 
\begin{itemize} 
    \item \textbf{MinIO}: Used for storing large source files and raster data, providing scalable and secure object storage.
    \item \textbf{PostgreSQL}: Manages all structured data, including metadata and vector data, to ensure data integrity through robust relational database management systems.
\end{itemize} 

\subsubsection{Backup and Recovery} 
\begin{itemize}
    \item Implement regular backup schedules for all data stored in MinIO and PostgreSQL.
    \item Maintain off-server backups for critical data to ensure recovery in case of major incidents.
    \item \textbf{Automated Orchestration and Logging}: Utilizing Prefect, automated capture, logging, and storage of metadata during data ingestion provides centralized monitoring and real-time tracking.
\end{itemize} 

\subsubsection{Metadata Capture} 
\begin{itemize} 
    \item Prefect is integrated with C-HER tools (e.g., Python) to provide dynamic metadata extraction and ensure robust error handling, making metadata documentation both reliable and scalable.
    \item To ease use and adoption, additional tooling was created for manually inserting and semi-automatically producing metadata about datasets.
\end{itemize} 

\subsubsection{Data Security} 
\begin{itemize} 
    \item Use encryption for data at rest and in transit.
    \item Implement strict access controls and authentication measures to safeguard data.
\end{itemize} 

\subsection{Data Access and Sharing} 

\subsubsection{Access Policies} 
\begin{itemize} 
    \item Define clear guidelines for internal and external data access.
    \item Use role-based access controls to restrict sensitive data exposure.
\end{itemize} 

\subsubsection{Data Sharing} 
\begin{itemize} 
    \item Facilitate data sharing with external collaborators through secure APIs and export functions.
    \item Ensure compliance with all legal and ethical standards for data sharing.
\end{itemize} 

\subsubsection{Data Quality Assurance} 
\begin{itemize} 
    \item Regularly validate data using automated tools and manual checks to ensure accuracy and reliability.
    \item Document all QA/QC procedures and maintain logs of all data checks.
    \item \textbf{Data Security}: User accounts allow access by schema.
\end{itemize} 

\subsubsection{Ethical Considerations} 
\begin{itemize} 
    \item Address privacy concerns, especially for data that involve personal information.
    \item Ensure that data usage complies with all applicable laws and ethical guidelines.
\end{itemize} 

\subsubsection{Data Archival and Disposal} 
\begin{itemize} 
    \item Archive data that is no longer actively used but has potential long-term value.
    \item Safely dispose of data that is no longer required, ensuring that all sensitive information is irretrievably deleted.
\end{itemize} 

\subsubsection{Compliance and Monitoring} 
\begin{itemize} 
    \item Regularly review data management practices to ensure compliance with this data management plan and relevant regulations.
    \item Adjust the plan as necessary to adapt to new technologies, data types, and compliance requirements.
\end{itemize}

\newpage

\section{Supplement 3. Supplementary Tables and Figures}
\label{sec:supplement-3}

\begin{table}[h]
\caption{Most important feature for each principal component (PC).
Council for the Advancement of Standards ID (CAS Standard ID) and chemical name.
An asterisk (*) indicates that the chemical is classified as IARC Group 1 or \textit{known carcinogen}.
Two asterisks (**) indicate that the chemical is classified as IARC Group 2 or \textit{probably a carcinogen}.}
    \label{tab:chemical_info}
    \vspace{1em}
    \centering
    \begin{tabular}{|c|c|c|}
        \hline
        PC & CAS Standard & Chemical Name \\
        \hline
        1 & 510-15-6 & Chlorobenzilate \\
        2 & 75-01-4 & Vinyl chloride* \\
        3 & 8001-35-2 & Toxaphene \\
        4 & 57-14-7 & 1,1-Dimethylhydrazine \\
        5 & 118-74-1 & Hexachlorobenzene \\
        6 & 100-44-7 & Benzyl chloride \\
        7 & 75-25-2 & Bromoform (Tribromomethane) \\
        8 & 122-34-9 & Simazine \\
        9 & 621-64-7 & N-Nitrosodi-n-propylamine \\
        10 & 302-01-2 & Hydrazine \\
        11 & 71-43-2 & Benzene* \\
        12 & 75-56-9 & Propylene oxide \\
        13 & 95-80-7 & 2,4-Diaminotoluene (2,4-Toluenediamine) \\
        14 & 52645-53-1 & Permethrin \\
        15 & N096 & Cobalt compounds \\
        16 & 23564-05-8 & Thiophanate-methyl \\
        17 & N590 & Polycyclic aromatic compounds \\
        18 & 107-13-1 & Acrylonitrile \\
        19 & 106-89-8 & Epichlorohydrin** \\
        20 & 100-41-4 & Ethylbenzene \\
        21 & 191-24-2 & Benzo[g,h,i]perylene \\
        22 & 67-66-3 & Chloroform \\
        23 & 7440-48-4 & Cobalt \\
        24 & 123-91-1 & 1,4-Dioxane \\
        25 & 75-55-8 & Propyleneimine \\
        \hline
    \end{tabular}
\end{table}
\end{document}